%% file: main.tex
\documentclass[letterpaper,twocolumn,10pt]{article}
\usepackage{usenix}
\usepackage{fontawesome}

\usepackage{xspace}
\usepackage{graphicx}
\usepackage{subfigure}
\usepackage{pifont}
\usepackage{threeparttable,booktabs,multirow,lscape}
\usepackage{multirow}
\usepackage{adjustbox}
\usepackage{enumerate}
\usepackage{listings}
\usepackage{tabularx}
\usepackage[ruled,vlined]{algorithm2e}
\usepackage{amsmath}
\usepackage{environ}
\usepackage{tikz}
\usepackage{color}
\usepackage{xcolor}
\usepackage{enumitem}
\usepackage{rotating}
\usepackage{threeparttable}
\usepackage[affil-it]{authblk}
\NewEnviron{elaboration}{
\par
\begin{tikzpicture}
\node[rectangle,minimum width=0.45\textwidth] (m) {\begin{minipage}{0.45\textwidth}\BODY\end{minipage}};
\draw[dashed] (m.south west) rectangle (m.north east);
\end{tikzpicture}
}
\usepackage{tcolorbox}

\usepackage{soul}
\newcommand{\tang}[1]{\textcolor{brown}{[Tang: #1]}}

\newcommand{\lei}[1]{{\color{red} Lei: #1}}

\newcommand{\etc}{\emph{etc}\xspace}
\newcommand{\ie}{\emph{i.e.}\xspace}
\newcommand{\eg}{\emph{e.g.}\xspace}

\newcolumntype{R}[1]{>{\raggedleft\arraybackslash}p{#1}}
\newcolumntype{L}[1]{>{\raggedright\arraybackslash}p{#1}}

\newcommand{\tab}{\hspace*{1em}}
\newcommand{\code}[1]{{\fontfamily{cmtt}\fontseries{m}\fontshape{n}\selectfont\small{#1}}}

\newcommand{\target}{\textsc{Culpritware}}

\newcommand{\normal}{benign}
\newcommand{\Normal}{Benign}
\newcommand{\abnormal}{\target{}}
\newcommand{\Abnormal}{\target{}}
\newcommand{\ccrime}{cyber-crime}
\newcommand{\Ccrime}{Cyber-crime}

\newcommand{\trick}{tactic}
\newcommand{\Malware}{Malware}
\newcommand{\malware}{malware}

\AtBeginDocument{%
  \providecommand\BibTeX{{%
    \normalfont B\kern-0.5em{\scshape i\kern-0.25em b}\kern-0.8em\TeX}}}

\begin{document}

\title{Lifting The Grey Curtain: A First Look at the Ecosystem of \textsc{Culpritware}}

\author[1]{Zhuo Chen}
\author[1]{Lei Wu}
\author[1]{Jing Cheng}
\author[2]{Yubo Hu}
\author[1]{Yajin Zhou}
\author[3]{Zhushou Tang}
\author[3]{Yexuan Chen}
\author[2]{Jinku Li}
\author[1]{Kui Ren}
\affil[1]{Zhejiang University}
\affil[2]{Xidian University}
\affil[3]{PWNZEN infoTech Co.,LTD}

\maketitle

\input{body/000_abstract}

\input{body/001_intro}
\input{body/002_background}
\input{body/003_methodology} 
\input{body/004_characterize}   
\input{body/005_understand}

\input{body/006_discussion}
\input{body/007_relatedwork}

\input{body/008_conclusion}


\bibliographystyle{plain}
\bibliography{main}

\input{body/appendix}

\end{document}

%% file: body/000_abstract.tex
\begin{abstract}

Mobile apps are extensively involved in \ccrime{}s.
Some apps are \malware{} which compromise users' devices, while some others may lead to privacy leakage. Apart from them, there also exist apps which \textit{directly make profit from victims  through deceiving, threatening or other criminal actions}.
We name these apps as {\target{}}. They have become emerging threats in recent years.
However, the characteristics and the ecosystem of \target{} remain mysterious.

This paper takes the first step towards systematically studying \target{} and their ecosystem. 
Specifically, we first characterize \target{} by categorizing and comparing them with \normal{} apps and \malware{}.
The result shows that \target{} have unique features, \eg, the usage of \textit{app generators} ($25.27\%$) deviates from that of \normal{} apps ($5.08\%$) and \malware{} ($0.43\%$). Such a discrepancy can be used to distinguish \target{} from \normal{} apps and \malware{}.
Then we understand the structure of the ecosystem by revealing the four participating entities (\ie, \textit{developer}, \textit{agent}, \textit{operator} and \textit{reaper}) and the workflow.
After that, we further reveal the characteristics of the ecosystem by studying the participating entities.
Our investigation shows that the majority of \target{} (at least $52.08\%$) are propagated through social media rather than the official app markets, and most \target{} ($96\%$) \textit{indirectly} rely on the covert \textit{fourth-party} payment services to transfer the profits.
Our findings shed light on the ecosystem, and can facilitate the community and law enforcement authorities to mitigate the threats.
We will release the source code of our tools to engage the community.

\end{abstract}

%% file: body/001_intro.tex
\section{Introduction}

Cyber-crime is a pervasive and costly global issue. 
The losses caused by \ccrime{}s are increasing every year~\cite{ic3Report}.
In 2020, the reported loss of global \ccrime{}s exceeds \$4.1 billion. 
According to the Federal Trade Commission (FTC)~\cite{ftcReport}, the mediums that facilitate \ccrime{}s include email, website, advertisement, \etc.
Previous works have paid attention to these mediums and studied how to combat \ccrime{}s~\cite{almomani2013survey, srivastava2008credit, mansfield2014dark, kreibich2009spamcraft}.

In recent years, mobile apps are extensively involved in \ccrime{}s. 
Previous studies~\cite{androidPha, rieck2008learning, felt2011survey, milletary2012citadel, idika2007survey} mainly target \textit{malicious apps} (\ie, malware, such as \textit{backdoor} and \textit{trojan}) that compromise/damage victims' devices.
Some other studies~\cite{roundy2020many, razaghpanah2018apps} focus on apps that may lead to privacy leakage and abuse (\eg, \textit{creepware}), which can be used to launch interpersonal attacks rather than to make profits.

\begin{figure}[t]
\centering
    \subfigure[]{
    \begin{minipage}[htb]{0.18\textwidth}
        \label{fig:gambling}
        \centering
        \includegraphics[width=1in]{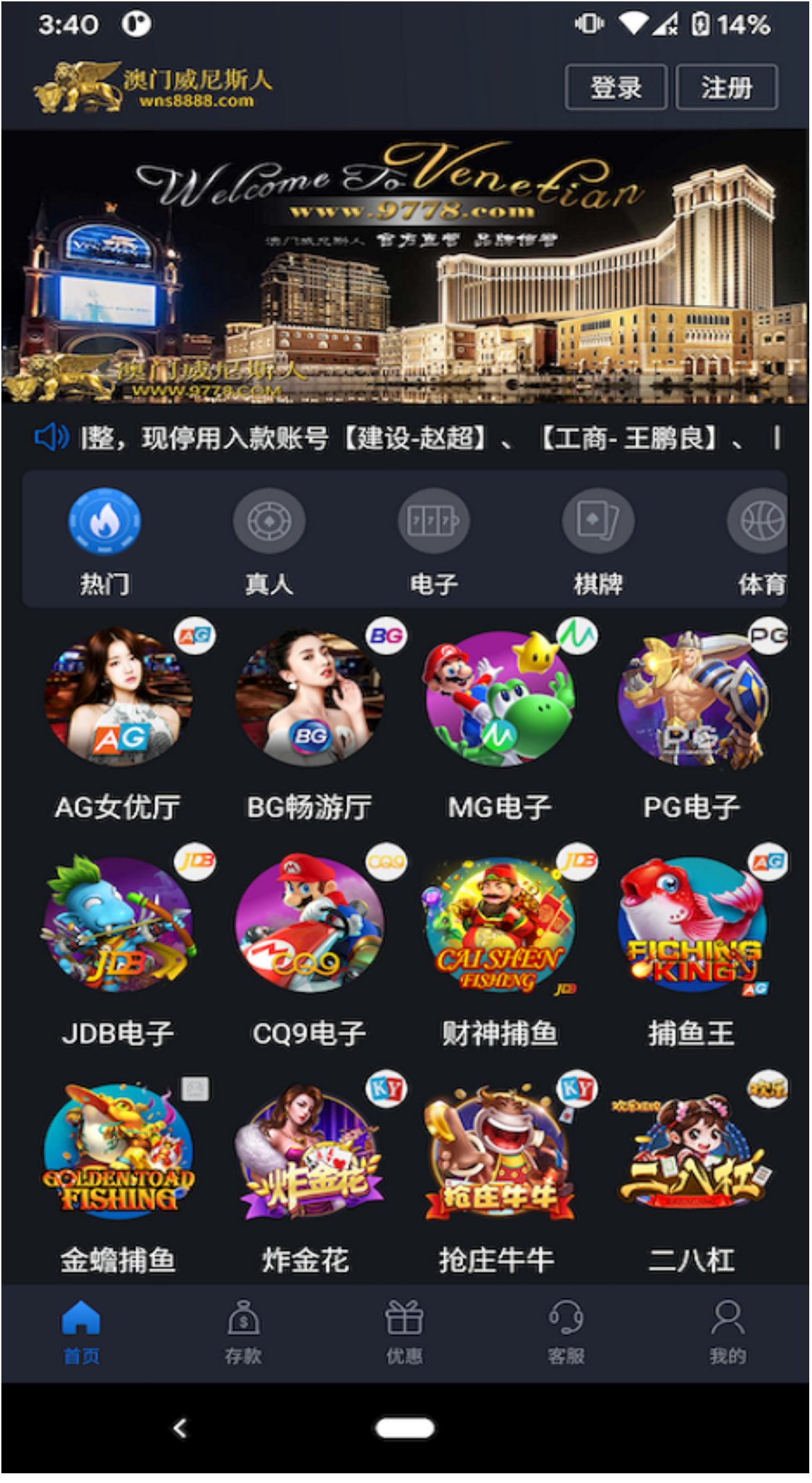}
    \end{minipage}}
    \subfigure[]{
    \begin{minipage}[htb]{0.18\textwidth}
       \label{fig:BTC}
       \centering
       \includegraphics[width=1in]{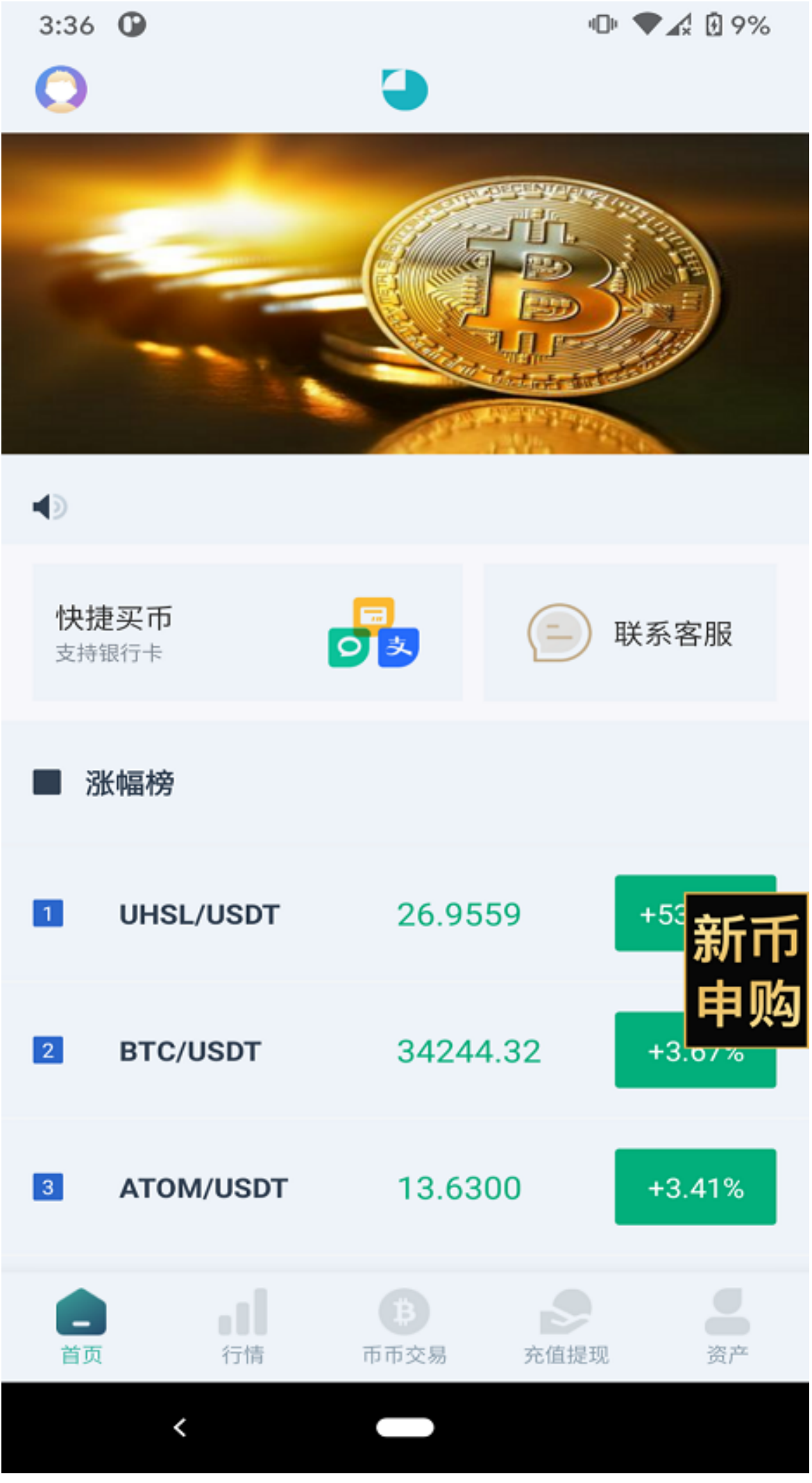}
    \end{minipage}}
 \caption{(a) A \textit{gambling} scam. (b) A \textit{financial fraud} app.}
 \label{fig:examples}
 \vspace{-0.2in}
\end{figure}

However, there do exist apps that do not fall into any of those categories.
These apps play an important role in reported attacks~\cite{muscanell2014weapons, atkins2013study, stajano2011understanding}, 
which \textit{profit from victims directly} through deceiving, threatening or other criminal acts, rather than compromising or damaging victims' devices.
Due to their unique illegal behaviors, we name these apps as \textit{culprit} apps, or \textbf{\target{}} for short. 
Figure~\ref{fig:examples} gives two examples. The first one is a gambling  scam app, while the second one is a financial fraud app, which pretends to be a cryptocurrency exchange but does not have this functionality.
We have witnessed the penetration of \target{} which already led to huge financial losses. 
For example, the pandemic has been leading to an increase in online shopping, which boosts the low-quality online shopping fraud. FTC estimated that this fraud has caused more than $\$245$ million in 2020~\cite{FTC}.
Another example is the romance scam that lures victims to invest on a dishonest investment app~\cite{romanceScammer}. Such scams have caused $\$304$ million losses in 2020~\cite{BBC}.
Besides, the ubiquitous gambling scam apps cause the highest financial losses in China.
According to a recent report published by the Chinese government~\cite{chinaGambling}, there are more than $3,500$ cross-border gambling and related cases in 2020.
As such, there is an urgent need for the security community to understand the characteristics of \target{} and its ecosystem.

Though some ad-hoc studies have been proposed to focus on specific domains of \target{}, \ie, gambling~\cite{binde2017forms} or dating scam~\cite{hu2018dating, suarez2019automatically, huang2015quit},
the characteristics of \target{} and the ecosystem remain mysterious.
A better understanding of \target{} and their underlying ecosystem is necessary and helpful to facilitate 
the mitigation of the threats.

This paper takes the first step towards systematically studying \target{}, including their characteristics and the underlying ecosystem.
To this end, we first establish three datasets, including \textit{\normal{}} dataset collected from reliable sources (\eg, Google Play Store~\cite{googleplay}), \textit{\malware{}} dataset collected from Virusshare~\cite{Virusshare}  and \textit{\target{}} dataset provided by an authoritative department.
In total, the \textit{\normal{}} dataset contains $90, 611$ apps, the \textit{\malware{}} dataset contains $1, 403$ apps and the \target{} dataset contains $843$ apps spanning from December 1, 2020 to June 1, 2021.

To demystify the \target{} and the ecosystem, we aim to answer the following three research questions. 
First, \textit{what are the characteristics of \target{}}? 
To characterize the \target{}, we first inspect the samples and establish the categorization criteria to reveal the \ccrime{} distribution (Section~\ref{subsec:categorize}).
After that, by comparing the \target{} with \normal{} apps and \malware{}, we reveal their development features  (Section~\ref{subsec:characterize}).
Second, \textit{what is the structure of \target{} ecosystem}?
To understand the ecosystem, we reveal the participating entities and the workflow of the \target{} ecosystem (Section~\ref{subsec:structurer_ecosystem}).
Third, \textit{what are the characteristics of \target{} ecosystem}?
To further reveal the characteristics of the ecosystem, we study the participating entities from the following aspects: the provenance of the developers (Section~\ref{subsec:developer}), the propagation methods (Section~\ref{subsec:propagation}), the management of the remote servers (Section~\ref{subsec:server}) and the covert payment services (Section~\ref{subsec:paymentservices}).
    
To the best of our knowledge, this is the first systematic effort to study \target{} and the ecosystem at scale, longitudinally, and from multiple perspectives. 
Our investigation provides a number of interesting findings, and the following are prominent:
\begin{itemize}[leftmargin=*]

\item { \textbf{\target{} have unique features, making them significantly different from \normal{} apps and malware.}}
Compared with \normal{} apps and \malware{}, \target{}'s developers are more inclined to develop hybrid apps, and abuse app generators to facilitate the development.
Specifically, the usage of \textit{app generators} ($25.27\%$) deviates from that of \normal{} apps ($5.08\%$) and \malware{} ($0.43\%$).

\item { \textbf{The \target{} ecosystem has formed a complete industrial chain.}} 
The ecosystem consists of four phases associated with four participating entities, \ie, \textit{developer} in the \textit{development} phase, \textit{agent} in the \textit{propagation} phase, \textit{operator} in the \textit{interaction} phase and \textit{profit reaper (reaper for short)} in the \textit{monetization} phase.

\item { \textbf{\target{} leverage covert distribution channels and use ``access code'' widely.}}
The majority of \target{} (at least $52.08\%$) are propagated through social media rather than app markets. Meanwhile, the ``access code'' is widely used (28.40\%) in \target, especially in the \textit{Sex} category, to make the propagation stealthy. A victim can even unwittingly become the conspirator
when the victim is lured inviting other users to the app.

\item { \textbf{The covert fourth-party payment services are indirectly abused to transfer the profits.}}
Most \target{} ($96\%$) adopt the fourth-party payments~\cite{forthParty},
which integrate multiple payment channels to provide covert payment services.
Specifically, they may rely on the third-party payments ($65.96\%$), the bank transactions ($23.40\%$) and the digital-currency payments ($8.51\%$), respectively. 
\end{itemize}

%% file: body/002_background.tex
\section{Background}

\subsection{App Development Paradigms}
\label{subsec:development}
There are three development paradigms, as follows:
\begin{itemize}[leftmargin=*]
    \item \textbf{Native apps: } 
    are developed in a platform-specific programming language (\eg, Android apps are developed primarily in Java). 
    They are not cross-platform compatible, but have better performance.
    \item \textbf{Web apps: } 
    are developed in web techniques (\eg, HTML, CSS, and JavaScript), which can be loaded in browsers (\eg, Chrome, Safari, Firefox) or embedded WebViews. They are not standalone, and must be as a part of other apps.
    \item \textbf{Hybrid apps: } 
    are the combination of the native apps and web apps. Hybrid apps are standalone like native ones, but internally they are built using web techniques~\cite{ali2016mining} and are most cross-platform compatible. 
    Hybrid apps can be separated into two parts: local clients and remote services.
    Local clients bridge user's requests to remote services, while remote services provide service per user's requests. 
\end{itemize}
In this paper, we only focus on the native apps and hybrid apps in the Android platform.

\subsection{App Generators}
\label{sec:app_generator}

The thriving \textit{app generators}~\cite{oltrogge2018rise} ease the app development process. 
App generators lower the level of technical skills required for app development by integrating the user supplied code snippet, media resource files or website addresses with the predefined template codes 
to build an app. What’s more, app generators try to simplify the pipeline of app development, distribution, and maintenance~\cite{SP2018appgenerator}, \eg, most app generators offer some encryption methods (\eg, RC4, TEA) to protect the entire APK file, while some online app generators (OAG) provide services to maintain app version upgrades.  All these features of app generators allow creating apps in batches.

\subsection{Covert Payment Services}
\label{subsec:payment_bg}
Rather than contacting victims directly, \target{} usually outsource the currency receiving process to the \textit{fourth-party} payment services~\cite{forthParty}. In general, the fourth-party payment can be built on the third-party payment~\cite{yang2017show}, the bank transaction or the digital-currency (\eg, Bitcoin and Litecoin)~\cite{brunt2017booted, cryptocurrencyScam}. 
Unlike formal payment services, the fourth-party payment services provide revenue transfer capabilities that evade supervision.
To transfer the illegal profits, the fourth-party payment services recruit workers, namely \textit{money mule}s~\cite{aston2009preliminary}, to bypass tracing. 
To provide a covert service, the fourth-party payment services use different recipient accounts to serve different requests. 

%% file: body/003_methodology.tex
\section{Study Design \& Data Collection}
\label{sec:method}

This study aims to demystify \target{} and their ecosystem by conducting comprehensive research. 
We seek to focus on the following three research questions (RQs):
\begin{itemize}
    \item[RQ1] \noindent \textbf{What are the characteristics of \target{}? } 
    No previous work characterizes \target{} in a comprehensive manner.
    By doing so, we can figure out abnormal behaviors and features to promote the identification and supervision of these apps. 
    
    \item[RQ2] \noindent \textbf{What is the  structure of \target{} ecosystem? } 
    We identify the participating entities and the workflow of the ecosystem to preliminarily understand the proliferation of \target{}.
    
    \item[RQ3] \textbf{What are the characteristics of \target{} ecosystem? } 
    We characterize the participating entities of the ecosystem to further delve into more details.
    The findings may be helpful in protecting the users against \target{} threats. 
\end{itemize}

\subsection{Our Approach}
\label{subsec:approach}

First, we answer RQ1 in Section~\ref{sec:characterizing}. 
We first manually categorize \target{}.
Based on that, we conduct an analysis on development features of \target{} on the development paradigm, use of app generators and the required permissions to reveal the differences among \normal{}, \malware{} and \target{} apps. 
Second, we answer RQ2 in Section~\ref{subsec:structurer_ecosystem} by revealing the participating entities of the ecosystem and the workflow.
Finally, we answer RQ3 in Section~\ref{subsec:characterize_ecosystem}. 
Our study is driven by four sub-questions about the developers, propagation methods, remote servers and payment services. 

\subsection{Data Collection}
\label{subsec:dataset}

As mentioned earlier, we need to prepare three datasets, \ie, \textit{\normal{}} dataset , \textit{\malware{}} dataset and \textit{\abnormal{}} dataset, to support our analysis:
\begin{itemize} [leftmargin=*]
    \item \textbf{\Normal{} Dataset: }
        apps in \textit{\normal{}} dataset are collected from reliable sources (\eg, Google Play~\cite{googleplay} and HUAWEI gallery~\cite{Huaweigallery}) from 2012 to 2021. In total, $90, 611$ android apps are  randomly chosen as the \normal{} dataset.
    \item \textbf{ \Malware{} Dataset: }
        apps in \textit{\malware{}} dataset are collected from Virusshare~\cite{Virusshare} (a virus sharing site) in 2018. In total, $1,403$ android malware are randomly chosen as the \malware{} dataset.  

    \item \textbf{\Abnormal{} Dataset: }
        apps in \textit{\abnormal{}} dataset are collected from an authoritative department from December 1, 2020 to June 1, 2021. 
        We get $1,076$ raw \target{} samples in total. 
        However, not all of the raw samples can be used for further analysis. For example, some of them are not valid for installation, which is important to determine their behaviors.
        Thus we need to remove these samples to guarantee the analysis.
        Finally, we make the \abnormal{} dataset with $843$ valid samples.

\end{itemize}
All apps in three datasets are sanitized by the package name and the checksum of their instances.
By comparing the difference among them, we are able to highlight the features of \target{}.

\smallskip 
\noindent \textbf{Ethics and Data Privacy}\tab 
The \target{} samples are delivered to authority from the victims by their own initiative. The dataset is authorized by an authoritative department without using any method that violates users' privacy or ethical concerns. Therefore, our dataset is guaranteed to be legal and authoritative.
The \target{} samples used in this paper were all involved financial or violent cases which were filed by a security expert. 

%% file: body/004_characterize.tex
\section{The Features of \target{}} 
\label{sec:characterizing}

To answer RQ1, we first categorize \target{} to reveal their different types and profit \trick{}s. 
Then, we characterize the development features of \target{} by comparing them with the \normal{} apps and \malware{}.

\subsection{Categorizing \target{}}
\label{subsec:categorize}

\subsubsection{Our Approach to Categorization}
In order to have an in-depth understanding, we first inspect \target{} samples to establish a categorization criteria.
We propose a three-step method to build the criteria, and
three security experts of our team work together as a group to implement the method accordingly:
\begin{itemize} [leftmargin=*]
	\item \textbf{Step 1: } 
    We randomly choose $200$ apps as the initial candidate set. 
    Each member of the group works independently to collect behavioral information (\eg, deposit procedure and user interaction) by manually investigating each sample. 
    \item \textbf{Step 2: }
    Based on collected information, the group establishes the consensus to assign an appropriate category name for each sample. 
    Meanwhile, the \trick{}s to make profits exploited by the samples are recorded.
    In this step, a categorization criteria matrix is established, including categories and their \trick{}s to make profits.
    \item \textbf{Step 3: } 
    Based on the matrix in the previous step, the remaining \target{} samples are added to the candidate set and further tagged by the group. 
    For each top category,
    its corner cases are labeled as ``miscellany'' (\eg, \textit{Sex Miscellany}) if the detailed information cannot be further identified.
    If there exists any bias, we return to the previous step by either refining improper categories or adding new categories. 
    
\end{itemize}
After applying the method, the criteria matrix is finalized in Table~\ref{tab:App_category} in Appendix.
The first column of the table shows the top-categories, the second column shows sub-categories of each top-category, and the last column shows the \trick{}s to make profits.
In total, we identify $5$ top-categories, $18$ sub-categories and $11$ \trick{}s. The \trick{}s are also detailed in Appendix. All samples are then tagged accordingly.

To better illustrate the categorization process, we present the process of inspecting a sample.
By investigating the app, we conclude the behavior as follows:

\begin{quote}\textit{
    ``This app pretends to be an official financial app and sells insurance products. 
    Specifically, this app requires credit card and other personal information (\eg, user name, ID card, telephone) and sends the information message to the remote server (December 27,2020).
    This app cannot be opened anymore (January 15, 2021).''}
    \vspace{0em}
\end{quote}
After step 1, our team members established the consensus to assign the category name as \textit{Financial} - \textit{Fake Financial Products} in step 2. 
However, in step 3, we find there exists bias in the \textit{Financial} - \textit{Fake Financial Products} category. 
Compared with other fake financial products, insurance-related apps require victims to provide their credit cards as a prerequisite. 
Typically, we show another \textit{Financial} - \textit{Fake Financial Products} app's \trick{} as a comparison:
\begin{quote}
    \vspace{0em}
    \textit{
    ``\dots only requires a telephone number and even without a format check or SMS verification code \dots after purchasing the financial product, the withdrawal function is unavailable (January 22, 2021)."}
    \vspace{-1em}
\end{quote}
To tackle the bias, we return back to step 2 and decide to divide the \textit{Financial} - \textit{Fake Financial Products} into two new categories named \textit{Financial - Insurance Products} and \textit{Financial - Financial Investment}. 
We use this iterative method to improve our categorization accuracy.

\smallskip 
\subsubsection{Categorization Results}
\label{subsubsec:cate_result}
To better understand the characteristics of these categories, we aggregate the common three usage phases of \target{}:
\begin{itemize} [leftmargin=*]
    \item  \textbf{User Seducement. } The first phase of \target{} is to entice the users. There are three different manifestations: (U1) fake information, (U2) disguise and (U3) free trial. 
    \item \noindent \textbf{Purchase/Deposit. } After app delivery, the \target{} focus on making revenue, they leverage multiple methods to get profit. We conclude their behaviors into three manifestations: (D1) activation fee, (D2) product/service and (D3) token purchase. 
    \item  \textbf{Follow-up Operation. }  After usage, the \target{} get money/sensitive information from the victims, their follow-up operations expose their crime identity. We also present three manifestations: (F1) none or low-quality delivery, (F2) disable functionality and (F3) contact interrupted. 
\end{itemize}

\begin{figure}[!t]
	\centering
	\includegraphics[width=.5\textwidth]{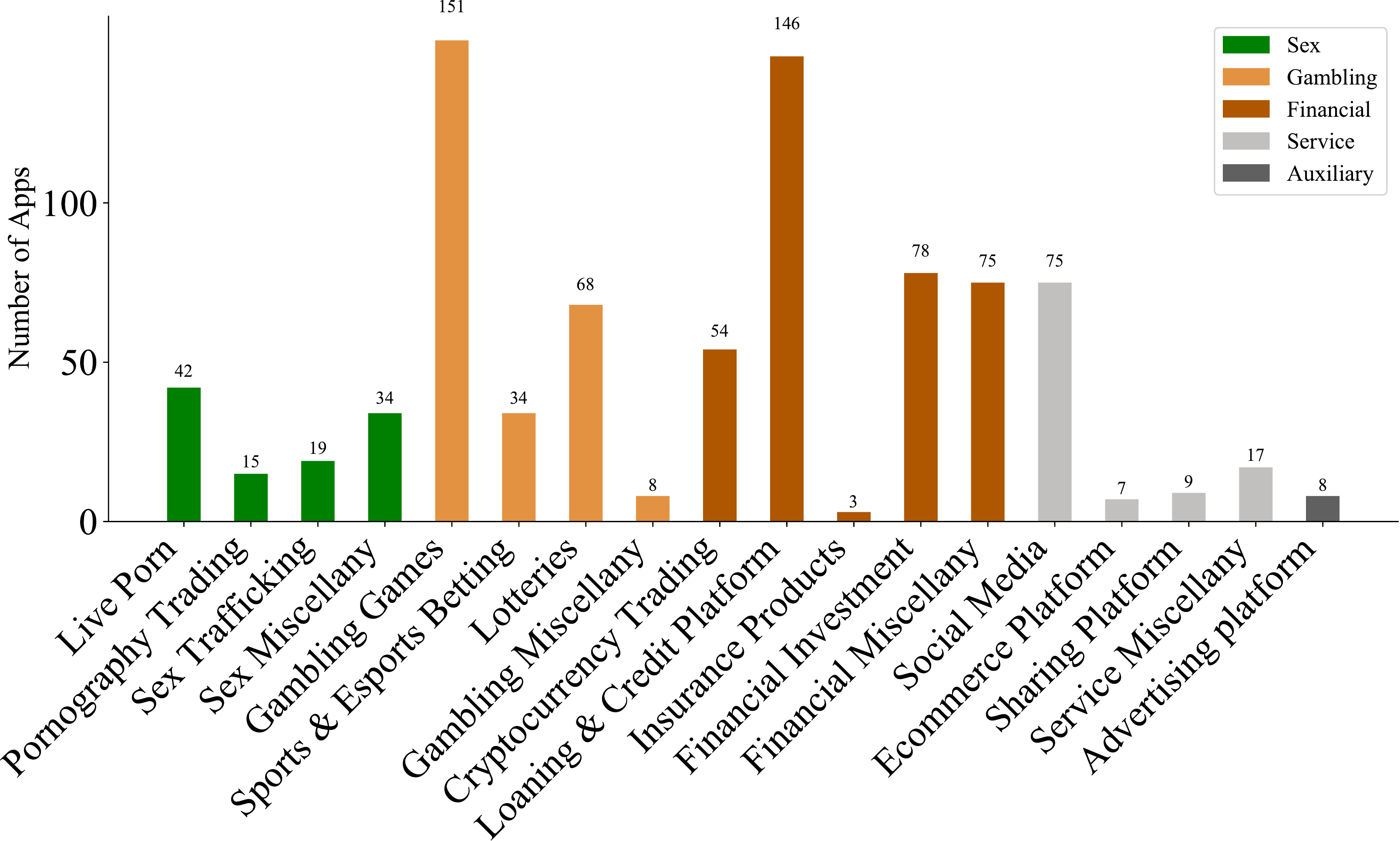}
	\caption{Categorization distribution of \target{}.}
	\label{fig:classification_result}
\vspace{-1em}
\end{figure}

The detailed behavior manifestations for each category are summarized in Table~\ref{tab:App_category} in Appendix. 
It shows that more than half ($51.37\%$) of the samples use false information to lure victims.
Specifically, in the \textit{Financial} top-category, they not only leverage false information, but also pretend to be regular software to confuse the victims. 
\target{} in different categories leverage different methods to harvest money from the victims and finally use disable functionality ($37.60\%$), contact interrupted ($16.13\%$) or low-quality delivery ($30.61\%$) to complete the crime. 
We conclude that the majority of \target{} take advantage of victims' greed, entice them to spend money of their own accord.

\smallskip 
\noindent \textbf{Distribution of the Categorization}.
The distribution of these categories is depicted in Figure~\ref{fig:classification_result}. 
The \textit{Financial} \target{} account for the largest portion, which make up $42.24\%$ of all \target{}. The second largest is the \textit{Gambling} which occupies $30.96\%$, followed by the \textit{Sex} ($13.04\%$), \textit{Service} ($12.82\%$) and \textit{Auxiliary Tool} ($0.95\%$).

\begin{tcolorbox}[size=title]
\textbf{Finding \#1:}
\target{} have high diversity of behavior categories associating with \trick{}s to make illicit profits, including 5 top-categories and 18 sub-categories with 11 profit \trick{}s. Among these categories, the \textit{Financial} ($42.24\%$) and \textit{Gambling} ($30.96\%$) \ccrime{} are the most prevalent ones.
\end{tcolorbox}

\subsection{Characterizing \target{}}
\label{subsec:characterize}

\subsubsection{Parse \target{}}
\label{subsubsec:observation}

Characterizing \target{} requires analyzing two types of apps, \ie, \textit{generator-based} apps and \textit{non generator-based} apps (see Section~\ref{sec:app_generator}), which need to be handled in different manners. 
On one side, for non generator-based apps, we can rely on existing tools like Androguard~\cite{desnos2011androguard} to gather information from the APK files.
On the other side, for generator-based apps, traditional tools have encountered some problems: 1) generator-based apps are glued with the template code provided by the app generators; and 2) some app generators encrypt the APK file to prevent reverse engineering.

As a result, it is necessary to decrypt generator-based apps if needed, and extract user-provided contents from the template code. Unfortunately, to the best of our knowledge, there is no effective analyzing tool designed for the generator-based apps.
To analyze generator-based apps, we investigate the app generators and gather the following observations:
\begin{itemize} [leftmargin=*]
    \item \textbf{Observation I:}
    There exist distinct features for each app generator, which can be used to distinguish the generator-based apps and the non generator-based ones. For example, for the apps generated by DCloud~\cite{Dcloud} (a major app generator provider), the main activity name in the manifest file is set to \verb|io.dcloud.PandoraEntry|.
    \item \textbf{Observation II:}
    The encryption method used by an app generator rarely changes for compatibility, which can be inspected and used to decrypt the resources.
    For example, AppCan~\cite{Appcan} uses native RC4 encryption and Appmachine~\cite{Appmachine} uses TEA encryption. 
    The app generators and the corresponding encryption methods are summarized in Table~\ref{tab:App_generator_collection} in Appendix.
    \item \textbf{Observation III:}
    The template code and resources offered by app generators mostly have a fixed structure, which can be used to distinguish the user-provided and generator-provided code and resources.
    Therefore, we can precisely extract user-provided code and resources from the generator-based apps.
\end{itemize}
Based on these observations, we build a tool (covering 47 kinds of app generators listed in Table~\ref{tab:App_generator_collection}) to automatically analyze generator-based apps.

To sum up, we combine all the tools together and apply them to extract the necessary data from those \target{} samples. The extracted data is organized and finally stored into the database for further analysis.

\subsubsection{Characterize \target{}}
\label{subsubsec:characterize_features}
Based on the data collected in the previous section, we find that there exist three distinct features, \ie, development paradigm, app generator usage and permission requirement,
which can be potentially used to detect \target{}.

\smallskip
\noindent \textbf{Feature I: development paradigm}.
We find $440$ hybrid apps, which occupies $52.19\%$ of all \target{}. 
While for the \normal{} and \malware{} datasets, the number (proportion) of hybrid apps are $14,851$ ($16.39\%$) and $114$ ($8.13\%$), respectively.
This huge difference shows that \textit{\target{} prefer the development paradigm of hybrid apps}. 

\smallskip
\noindent \textbf{Feature II: app generator usage}.
To verify the influence of app generators in the development process of \target{}, we detect generator-based apps in our dataset.
From $843$ apps in the \target{} dataset, 
we identify $7$ different app generators in $213$ apps, and the proportion of the generator-based apps is $25.27\%$ of all apps.
While for the \normal{} and \malware{} datasets, the proportion of the generator-based apps are only $5.08\%$ and $0.43\%$, respectively. 
\begin{figure}[!t]
    \centering
    \includegraphics[width=.45\textwidth]{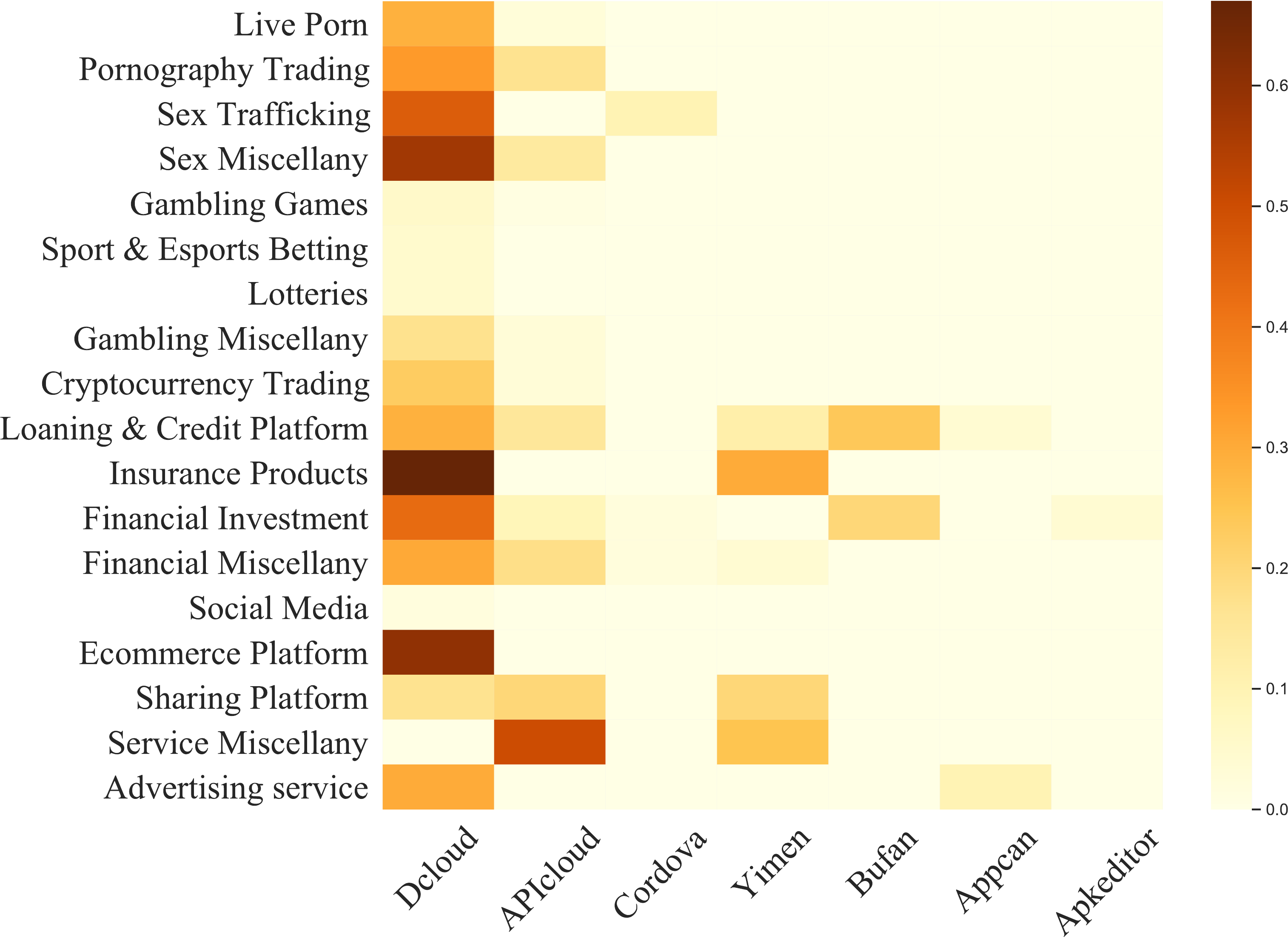}
        \caption{Distribution of app generators in \target{}.}
    \label{fig:APP Generators}
\vspace{-1em}
\end{figure}

Figure~\ref{fig:APP Generators} gives the distribution of app generators for different sub-categories.
The result shows that the apps in top-category \textit{Financial} use app generators ($43.82\%$ of all \textit{Financial} apps) more frequently than the others.
Even in the least frequently used category, \ie, the gambling, the rate is still $11.12\%$.
Alternatively, the most frequently used app generators are \verb|DCloud| and \verb|APICloud|, occupying $69.01\%$ and $18.78\%$, respectively; while others (\eg, \verb|Appcan| and \verb|apkeditor|) are rarely used.

As such, we conclude that \textit{app generators have been extensively abused to facilitate the development of \target{}}.
Namely, these app generators need additional regulation and supervision to prevent the proliferation of \target{}.

\smallskip
\noindent\textbf{Feature III: distribution of app permissions.} 
We aim to study (dangerous) permissions used by \target{}. 
Due to the security design of the Android system, apps need to be explicitly authorized  by users to obtain access to dangerous permissions~\cite{googlepermission}, such as location information (\verb|ACCESS_FINE_LOCATION|) and camera access (\verb|CAMERA|).
The acquired dangerous permissions are an important indicator to measure the intention and capability of the app~\cite{arora2019permpair}. 

\input{tables/permission}

For each \target{} sample, we examine all permissions from the manifest file and identify dangerous permissions defined by Google~\cite{googlepermission}. 
Table~\ref{tab:permissions} lists the result, which shows that the permission requirements of \target{} are more than \normal{} apps, but fewer than \malware{}. 
The average number of dangerous permissions is  $3.96$ for \target{}, while the numbers of \normal{} apps and \malware{} are $1.36$ and $4.53$, respectively.
Specifically, the majority of \target{} require $3$ or $4$ dangerous permissions, while some of them even require $15$ dangerous permissions.
Alternatively,
the average number of normal permissions is $19.36$ for \target{}, while the numbers of \normal{} apps and \malware{} are $6.02$ and $29.82$, respectively.
To sum up, \textit{the average number of permissions used by \target{} ($23.32$) deviates from that of \normal{} apps ($7.38$) and \malware{} ($34.35$)}.

\begin{tcolorbox}[size=title]
\textbf{Finding \#2:}
\target{} own unique features, making it significantly different from \normal{} apps and \malware{}.
Compared with \normal{} apps and \malware{}, \target{}'s developers tend to develop hybrid apps, and abuse app generators to facilitate the development.
Besides, \target{} use more permissions than the \normal{} apps, while less permissions than the \malware{}.
\end{tcolorbox}

%% file: tables/permission.tex
\begin{table}[t]
\centering
\footnotesize
\caption{Permission requirement analysis.}
\label{tab:permissions}
\begin{threeparttable}
\begin{tabular}{cllll}
\toprule
\multicolumn{1}{l}{Type}   & Category & Dangerous & Normal  & All \\
\midrule
\multicolumn{1}{c}{} &Sex  & 4.90  & 20.57 & 25.47   \\
\multicolumn{1}{c}{} & Gambling &  3.79 & 22.06 &  25.85    \\
\multicolumn{1}{c}{} & Financial  & 3.49 & 15.16 &  18.65   \\
\multicolumn{1}{c}{} & Service & 6.18 & 42.61 & 48.79     \\
\multicolumn{1}{c}{} & Auxiliary Tool & 4.50 & 33.17 & 37.67    \\
\cmidrule(lr){2-5}
\multicolumn{1}{c}{\multirow{-7}{*}{Culpritware}} &  \textbf{Total} & \textbf{3.96} & \textbf{19.36} & \textbf{23.32}\\
\midrule
\multicolumn{1}{l}{Malware{}} & \textbf{Total} & \textbf{4.53} & \textbf{29.82} & \textbf{34.35}
\\
\midrule
\multicolumn{1}{l}{Benign} & \textbf{Total} & \textbf{1.36} & \textbf{6.02} & \textbf{7.38}
\\
\bottomrule
\end{tabular}

\begin{tablenotes}
\footnotesize
\item[] The numbers in this table represent average number of permissions.
\end{tablenotes}
\end{threeparttable}

\vspace{-2em}
\end{table}

%% file: body/005_understand.tex
\section{The Ecosystem of \target{}}
\label{sec:understanding}
In this section, we aim to provide a comprehensive understanding of the \target{} ecosystem.
To this end, we first reveal the structure of the ecosystem to answer RQ2, and then characterize the ecosystem to answer RQ3.

\input{body/0051_structure}

\input{body/0052_characteristic}

\input{body/0053_developers}

\input{body/0054_propagation}
\input{body/0055_remote}

\input{body/0056_payment}

%% file: body/0051_structure.tex
\subsection{The Structure of the Ecosystem}
\label{subsec:structurer_ecosystem}
The \target{} ecosystem is a diverse and complex system with different participating entities. 
Our observation suggests that these entities have already formed an underground industrial chain to create, distribute and maintain \target{}.
Figure~\ref{fig:ecosystem} depicts the \target{} ecosystem, which is composed of the following four phases:

\begin{figure}
	\centering
	\includegraphics[width=.45\textwidth]{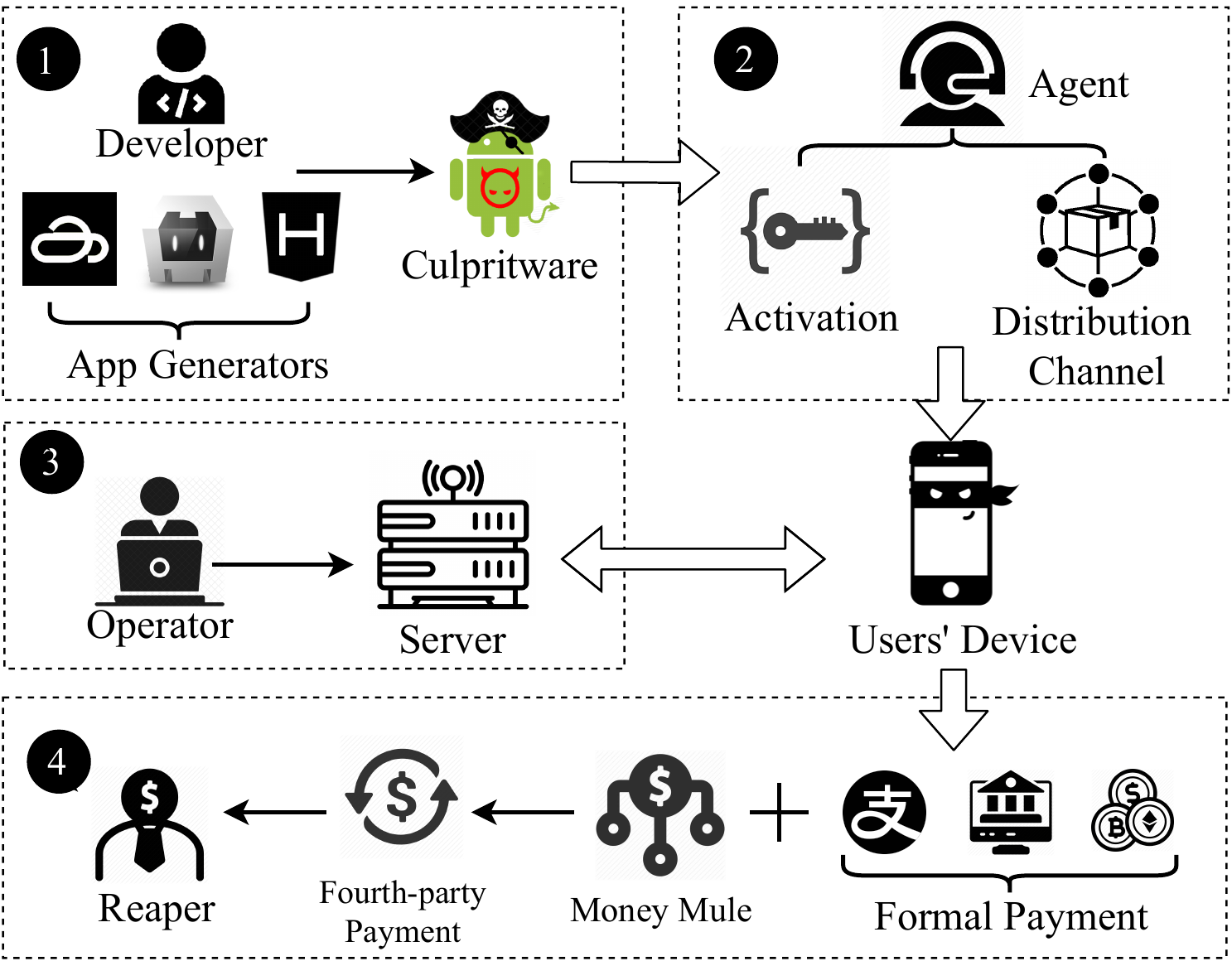}
	\caption{The \target{} ecosystem.}
	\label{fig:ecosystem}
\end{figure}

\begin{itemize} [leftmargin=*]
	\item \textbf{The Development Phase. } 
	In the first phase (\ding{182} in Figure~\ref{fig:ecosystem}), the \textit{developer} is responsible for creating \target{}.
    According to the features characterized in Section~\ref{subsubsec:characterize_features}, there exist two methods to develop \target{}, \ie, building the app directly, or leveraging the app generators to perform the task. 
    Our investigation suggests that the latter method has been abused widely in this ecosystem due to the following two reasons: 1) it does lower the bar and reduce the cost for the development; 2) it does not enforce strict supervision. 
     
	\item \textbf{The Propagation Phase. } 
    In the second phase (\ding{183} in Figure~\ref{fig:ecosystem}), the \textit{agent} distributes \target{} to users' devices, and to ``guide'' the users to activate/use \target{}. 
    Specifically, the agent may leverage some channels to distribute \target{}. Such channels can either be social media (apps)~\footnote{Such as chatting groups on legal platforms (\eg, \textit{Telegram}\cite{telegram} and \textit{WeChat}\cite{Wechat}) and illegal chatting/advertising/sharing platforms.}, websites containing the corresponding download links or traditional distribution channels (\eg, SMS, email and telephone). 
    After that, the agent should assist victims to activate \target{}, \ie, registering accounts. In many cases, \target{} may require the ``access codes'' to activate them. As a result, the activation process becomes a daily routine managed by the agent.
     Note that most of the users are \textit{victim}s, while others may become \textit{accomplices} (see Section~\ref{subsubsec:invite} for details).
    
    \item \textbf{The Interaction Phase. } 
     In the third phase (\ding{184} in Figure~\ref{fig:ecosystem}), the \textit{operator} will actively interact with victims to launch frauds/scams. The profit \trick{}s summarized in Section~\ref{subsec:categorize} indicate the interaction methods adopted by the operators for different \target{}, and we summarize these typical operators of the profit \trick{}s in Table~\ref{tab:operators} in Appendix. 
    For example, for the \textit{Romance Fraud} (``P2'' in the profit \trick{}s), there exist porn sellers, who specialize in communicating with victims and using romantic traps to make revenue from the victims. 
    Besides, the operator is also responsible for maintaining the operations of remote servers, which are used to store sensitive information (\eg, porn photos, gambling internal logic and transaction list). 
    This information is usually transferred to victims' devices upon their requests. 
    
    \item \textbf{The Monetization Phase. } 
    In the fourth phase (\ding{185} in Figure~\ref{fig:ecosystem}), the \textit{reaper} harvests the victims’ money. 
    Unsurprisingly, the reaper usually leverages the \textit{fourth-party} payment services
    to surreptitiously transfer profits. The fourth-party payment services are built on third-party payments, digital-currency payments or bank transactions. They also recruit money mules to bypass tracking (see Section~\ref{subsec:paymentservices}).
\end{itemize}
    
\begin{tcolorbox}[size=title]
\textbf{Finding \#3:}
The \target{} ecosystem is composed of four phases associated with four entities, \ie, the \textit{developer} in the \textit{development} phase, the \textit{agent} in the \textit{propagation} phase, the \textit{operator} in the \textit{interaction} phase and the \textit{reaper} in the \textit{monetization} phase.
\end{tcolorbox}

%% file: body/0052_characteristic.tex
\subsection{The Characteristics of the Ecosystem}
\label{subsec:characterize_ecosystem}
After revealing the structure of the ecosystem, we need to demystify more detailed characteristics to answer RQ3. Our study is driven by the following four sub-questions:
\begin{itemize}
    \item[RQ3-1] \textbf{Who are the developers of \target{}? } 
    Although the development of \target{} has been revealed in the previous sections, the developers behind remain unknown.
    To address this issue, we conduct the \textit{developer analysis} (in Section~\ref{subsec:developer}) to profile the developers of \target{} by first associating and then tracing the provenance for their real-world identities.
    
    \item[RQ3-2] \textbf{How do \target{} propagate among victims? } 
    We have roughly described the propagation phase in the previous section, however, the details still need to be demystified, including the distribution channels, the registration and invitation process. We perform the \textit{propagation analysis} (in Section~\ref{subsec:propagation}) to delve into the details.
    
    \item[RQ3-3] \textbf{How do the operators provide stable and stealthy services? } 
    Due to the improvement of regulatory enforcement (\eg, General Data Protection Regulation (GDPR)~\cite{GDPR}), it is necessary for the operators to hide the remote servers to provide stable and stealthy services. 
    To understand these behaviors, we launch the \textit{remote server analysis} (in Section~\ref{subsec:server}) to analyze the lifespan, the geographic location, the manipulation of domains and IP addresses, and the domain registrants.
    
    \item[RQ3-4] \textbf{What are the characteristics and usage of these payment services?} 
   Due to the strict law regulation, the formal payment services (\eg, the third-party payment) cannot be used by the reapers to transfer their illicit profits. Instead, they have to seek and use the covert payment services.
   As such, we present the \textit{payment analysis} (in Section~\ref{subsec:paymentservices}) to study the representative covert payment services leveraged by \target{}.
\end{itemize}

%% file: body/0053_developers.tex
\subsection{The Developers of \target{}}
\label{subsec:developer}
In this section, we aim to profile the developers behind \target{} to answer RQ3-1.
To this end, we first collect information that may be used to distinguish/correlate developers of \target{}, including the developer signatures, the URL lists and the GUI snapshots, to support further analysis.
Based on the collected data, we then conduct an association analysis to determine the possible relationship between developers.
Finally, we study the provenance of developers by associating \target{} with more apps, which may reveal the real-world identities of those developers.

\subsubsection{Pre-processing}
\label{sec:pre}
For all samples in the \abnormal{} dataset, we collect the following data:
\begin{itemize} [leftmargin=*]
	\item \textbf{Developer signatures. } 
	We first filter out public known signatures issued by some organizations (\eg, the Android Studio provided signatures~\cite{debug} and default signatures of app generators~\cite{SP2018appgenerator}), as these signatures do not provide any useful information for our analysis.
	Among these samples, $17.4\%$ are associated with public known signatures, \ie, $7.9\%$ are Android Studio signatures and $9.5\%$ are app generator signatures, respectively.  
	After that, we parse the signature fields to support the fine-grained analysis. 
    Note that $14.98\%$ fields are incomplete (or even blank), which need to be marked so as not to disturb the further analysis.
	By doing so, the qualified fields of signatures are collected. 

	\item \textbf{URL lists. } 
	We extract domains and IP addresses from the samples based on the fixed format (\eg, http/https format, IPv4/IPv6 format). 
	To improve the effectiveness, we also filter out common URLs by using Alexa~\cite{Alexa} index.

	\item \textbf{GUI snapshots. } 
    To collect snapshots of samples, we automatically install and launch the APK files on devices and take snapshots.
	Specifically, for a given app, we capture a snapshot every 15 seconds for the first minute after its startup and send all snapshots back to the database.
	Furthermore, for snapshot errors caused by permission popover, we customize a firmware image by modifying the source code of AOSP~\cite{AOSP} to grant all required permissions.
\end{itemize}

\subsubsection{Developer Association Analysis}
To trace the provenance of the developers, we propose a similarity-based approach to perform the developer association analysis in Algorithm~\ref{alg:cluster} in Appendix. 
Previous studies have demonstrated the effectiveness to associate developers by applying single-attribute (\eg, signature~\cite{sebastian2020towards}, GUI~\cite{pears2013weighted}) based approaches. We extend this idea to multiple attributes, including signature, URL list and GUI, to support the analysis.

\input{tables/association}

The algorithm accepts the entire sample set as the input, and iteratively performs association to generate a developer association graph. To prevent dependency explosion and infinite looping, we limit the maximum iteration count as $I_{max}$.
Specifically, the association rules for different type of attributes are defined as follows:
1) \textit{signature association} (\ding{182} in Algorithm~\ref{alg:cluster}): if the contents of the same field are the same, we associate the corresponding samples; and
2) \textit{URL association} (\ding{183} in Algorithm~\ref{alg:cluster}): we compare URL lists and associate them if the overlapped part is over 70\%~\footnote{We select this number after adjusting according to the association result.}. And also associate apps if their URLs have used the same IP ( see Section~\ref{subsubsec:manipulation}); and
3) \textit{snapshot association} (\ding{184} in Algorithm~\ref{alg:cluster}): we leverage SIFT algorithm~\cite{lowe2004distinctive} to extract key points and calculate Euclidean distance as the similarity of snapshot images.
Finally, the apps in the same associated group share the same developer attributions, hence the developers behind them are highly likely to be the same one. 
We apply this algorithm and find $287$ groups. 
The top 10 groups are summarized in Table~\ref{table:association}.

Based on the result, we identify several groups of samples with the large sizes.
The biggest group consists of $85$ apps ($10.08\%$ of all samples), then the second consists of $72$ apps ($8.54\%$).  
In conclusion, the majority of apps (74.73\% of the entire dataset) can be associated using the algorithm. Therefore, the majority of the \target{} can be identified by correlation with existing samples.
Besides, from the composition of the top 10 groups, we notice that the majority of apps in a group belong to the same category. Especially in the small groups, even all of them are consist of the same category. 
We then conclude that many illegal developers mass-produces \target{} of the same category.

In conclusion, we find that the constitution of developers behind the \target{} samples does not obey the well-known \textit{80/20 rule} (aka, \textit{Pareto principle}~\cite{pareto-principle}), which indicates the majority of \target{} are developed by individual developers who only develop $1$ or $2$ apps in the same category.
\begin{figure}[!t]
    \centering
    \includegraphics[width=.45\textwidth]{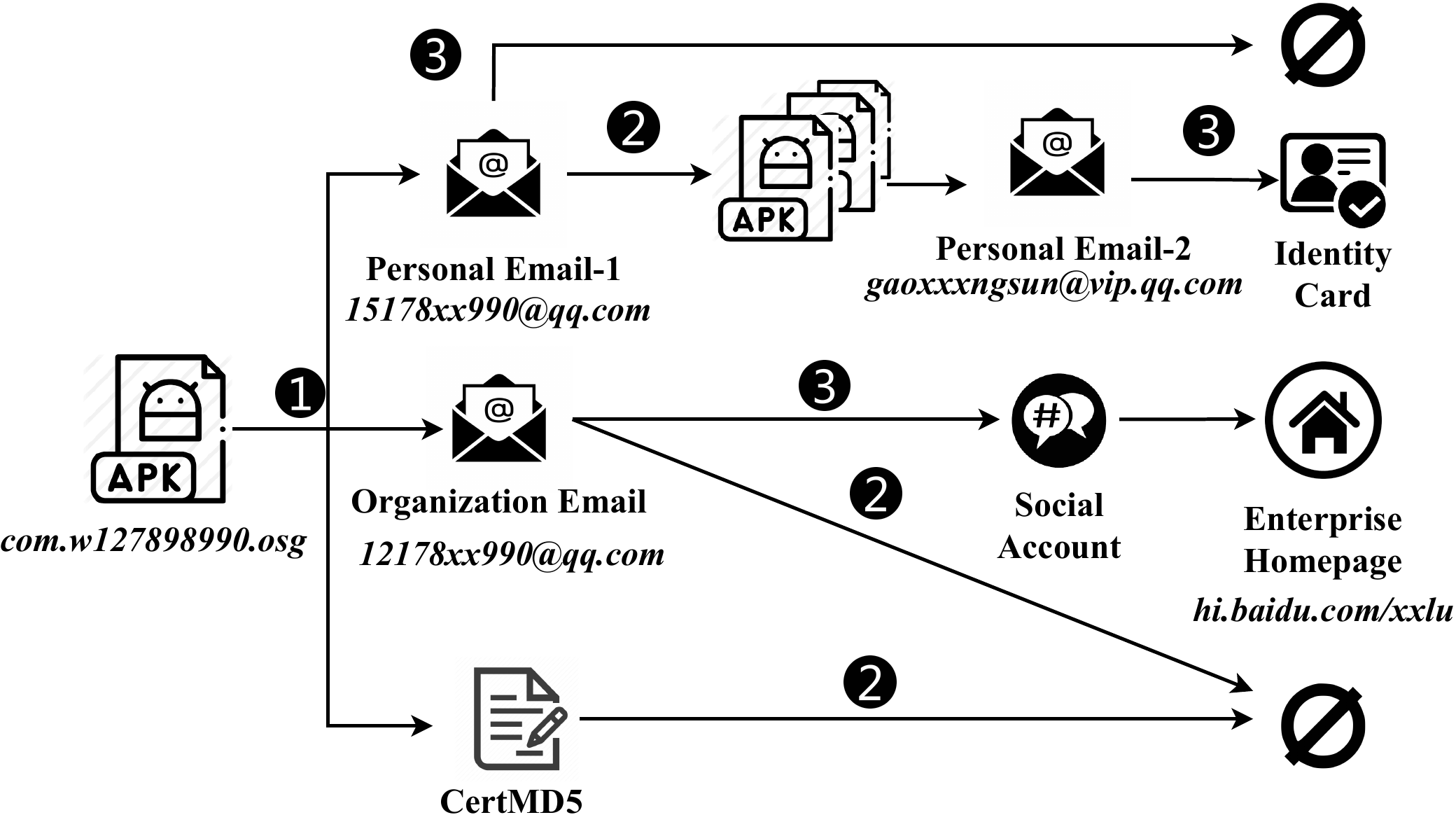}
    \caption{An example of developer provenance.}
    \label{fig:provenance}
\vspace{-1em}
\end{figure}

\subsubsection{Revealing the provenance of developers}
\label{subsubsec:provenance}
After the developer association process, we further conduct a provenance process of developers behind \target{}.
The method is summarized in the following three steps:
\begin{itemize} [leftmargin=*]
	\item \textbf{Summarizing information: } 
	\ding{182} in Figure~\ref{fig:provenance}. 
	We collect information from the apps in our dataset (\eg, the checksum of signature, email address, organization name). 
	
	\item \textbf{Expanding related apps: } 
	\ding{183} in Figure~\ref{fig:provenance}. 
	Considering that our dataset is limited, we use aggregated information to search for related apps in various Android application collections (e.g, Koodous~\cite{koodous}, XingYuan~\cite{xingyuan}) to expand our dataset. Through these associated apps, we obtain more potential features of the same developer.
	
	\item \textbf{Associating real identity: }
	\ding{184} in Figure~\ref{fig:provenance}. 
	We further use several cyberspace search engines (\eg, ZoomEye~\cite{ZoomEye}~\footnote{A search engine that lets the user locate specific network components.}, Tianyancha~\cite{Tianyancha}~\footnote{A service to query (Chinese) enterprise information and business data.}) and online forums to search for relevant information about the developer.
\end{itemize}
Targeting the top 10 app groups, we find $2$ enterprises in the Tianyancha~\cite{Tianyancha}, $5$ social accounts and $1$ ID card information.
We observe that there exist two types of developers:
1) specialized enterprises or studios engaged in app development; and
2) hired technicians to create customized apps.

To better illustrate our provenance process, we take an app \textit{com.w1217898990.osg} as an example and summarize the workflow in Figure~\ref{fig:provenance}.
After step \ding{182}, we get the email addresses  (personal email-I, organization email) and the checksum of signature.
Then, we perform step \ding{183} and find that the personal email associates $18$ apps through Koodous~\cite{koodous}. 
Through the associated apps, we collect the personal email-II.
Finally, after the step \ding{184}, we find an organization email, which correlates to a social account (Tecent QQ~\cite{QQ}), an enterprise homepage.

\begin{tcolorbox}[size=title]
\textbf{Finding \#4:}
The developer constitution behind \target{} does not seem to obey the \textit{80/20 rule}.
The provenance study reveals that some of the developers are registered companies while others are hired technicians.
\end{tcolorbox}

%% file: tables/association.tex
\begin{table}[!t]
\centering
\scriptsize
\caption{Top 10 groups of the developer association result\tnote{1}.}
\begin{threeparttable}
\begin{tabular}{cccccc}
\toprule
& & \multicolumn{4}{c}{Categories Composition} 
\cr\cmidrule(lr){3-6} 
Rank & Apps (\%Percent) & Sex & Gambling &Financial & Service \cr  
\midrule
1  & 85 (10.08\%)& 20 (23.5\%)  &  11 (12.9\%) &  52 (61.2\%) & 2 (2.4\%)   \\
2 &  72 (8.54\%) &  35 (48.6\%)&  0 & 37 (51.4\%) & 0    \\
3  & 27 (3.20\%)&  1 (3.7\%)   & 5 (18.5\%) & 21 (77.7\%)  & 0  \\
4  & 20 (2.37\%) &  3 (15.0\%) & 0  &17 (85.0\%) &  0   \\
5 & 15 (1.79\%) & 0 & 0 & 15 (100\%) & 0  \\
6 & 15 (1.79\%)& 0 & 15 (100\%) & 0 & 0  \\
7 & 13 (1.54\%) & 0 & 1 (7.7\%) & 9 (69.2\%)  & 3 (23.1\%)   \\
8 & 12 (1.42\%) &  11 (91.7\%) & 0 & 1 (8.3\%) & 0   \\
9 & 12 (1.42\%)& 1 (83.3\%) & 3 (25.0\%) & 4 (33.3\%) & 4 (33.3\%)  \\
10 & 10 (1.19\%) & 0 &  0 &  9 (90.0\%) &  1 (10.0\%)   \\
\bottomrule
\end{tabular}
\begin{tablenotes}
\footnotesize
\item[] Since the \textit{Auxiliary tool} does not appear in the top 10, we remove this category from the table.
\end{tablenotes}

\end{threeparttable}
 \label{table:association}
 
\vspace{-2em}
\end{table}

%% file: body/0054_propagation.tex
\subsection{The Propagation of \target{}}
\label{subsec:propagation}
In this section, we investigate the approach to propagate \target{} to answer RQ3-2.
Our analysis focuses on the following three perspectives, \ie, the distribution channels, the registration process, and the invitation process.

\subsubsection{The Distribution Channels}
As discussed in Section~\ref{subsec:structurer_ecosystem}, \target{} can be propagated through channels like social media (apps), websites containing the download links, and traditional distribution channels. 
Specifically, we analyze the distribution channels leveraged by \target{}. 
We find that the majority of the \target{} samples (at least $52.08\%$) are propagated through social media, while part of them (at least $5.21\%$) are downloaded through advertisement from other online websites (\eg, job hunting sites, cyber forum). Besides the online propagation, there are also a few apps (at least $7.29\%$) propagate through traditional distribution methods (\eg, SMS, Email or telephone).
Note that there are some apps ($35.42\%$) that lack the precise source of user acquisition.
Our experience suggests that most of them come from the three channels~\footnote{There also exist some channels that are rarely used, \eg, offline advertising and bluetooth transmission, we ignore them in this study.}, thus the actual proportion of these channels might be greater.

Meanwhile, we try to figure out whether \target{} are propagated through formal channels, \ie, the official app stores~\footnote{We focus on some well-known and representative app stores, including Google Play Store~\cite{googleplay}, HUAWEI AppGallery~\cite{Huaweigallery} and MI Store~\cite{Mistore}.}.
Namely, we need to verify the existence of \target{} in these app stores. 
To this end, we implement a tool which can monitor and record relevant changes of the app stores, \ie, all apps being added or removed for a time period.

We have continuously monitored these app stores for $6$ months, \ie, from December 1, 2020 to June 1, 2021, which is the same time period to collect the \abnormal{} dataset (Section~\ref{subsec:dataset}).
Interestingly, \textit{no \target{} samples are ever observed in any of the official app stores}. 
This observation suggests that agents are inclined to propagate \target{} through covert channels,
which is probably due to the regulation/supervision of the official channels.

\input{tables/Registration}
\subsubsection{The Registration Process}
In this section, we'd like to study the registration process of \target{} by comparing it with that of the \normal{} apps.
Specifically, we focus on two aspects, \ie, the way to register a new account and the use of End User Licence Agreement (EULA). The former is an important feature for the propagation, while the latter is related to the privacy issue.

To this end, we randomly collect 250 \target{} apps which are still active from the \abnormal{} dataset and 250 apps from the \normal{} dataset, respectively. 
We manually investigate and summarize the result in Table~\ref{tab:invite}. 
On one hand, many \target{} samples ($28.40\%$) require an ``access code'' as a prerequisite for registration, while no benign apps have the same requirement.
We also notice that the ``access code'' rate is extremely high ($55.26\%$) in the \textit{Sex} top-category, while the rate is much lower in other top-categories like \textit{Auxiliary ($0\%$)} and \textit{Service ($4.65\%$)}.
On the other hand, the absence of EULA is very common for \target{}, \ie, $59.60\%$ of the selected samples do not have a valid EULA. 
The proportion of the benign ones, by contrast, is only $3.20\%$.

The result suggests that \textit{the ``access code'' is widely used to control the propagation, thereby reducing the risk of exposure to the supervision; meanwhile, the agents are apt to target the victims who may lack security (and legal) awareness}.

\subsubsection{The Invitation Process}
\label{subsubsec:invite} 
Based on the previous analysis, we further find that the ``access code'' is associated with the invitation process, which may turn the user into a new agent, namely, an accomplice.
To reveal the details, we manually investigate the invitation process by communicating with the customer agents of \target{}. 
Most customer agents maintain high vigilance and refuse to answer our questions.
Fortunately, we still contact a few of them to acquire some useful information.

Specifically, users can participate in the propagation process using two methods.
First, sharing ``access code'' with acquaintances. 
\target{} may reward the distributors for their contributions with some small cash.
Second, joining the ecosystem as an agent.
The user has the opportunity to become an agent who is responsible for propagating \target{} in one (small) area. 
Obviously, these designs facilitate and boost the propagation of \target{}.

\begin{tcolorbox}[size=title]
\textbf{Finding \#5:}
The majority of \target{} (at least $52.08\%$) are propagated through social media rather than the formal app markets. Meanwhile, the ``access code'' is widely used (28.40\%) in \target, especially in the \textit{Sex} category. The users can even become the accomplices through the invitation process.
\end{tcolorbox}

%% file: tables/Registration.tex
\begin{table}[!t]
\caption{Registration and EULA analysis.}
\centering
\footnotesize

\begin{tabular}{cllllll}
\toprule
\multicolumn{3}{l}{{\color[HTML]{000000} Type}}   & {\color[HTML]{000000} Category} & Number & Access Code & EULA Absence \\
\midrule
\multicolumn{3}{c}{} & Sex & 38  & 21 (55.26\%) & 26 (68.42\%)    \\
\multicolumn{3}{c}{} & Gambling   &  77 & 17 (22.08\%) & 51 (66.23\%)      \\
\multicolumn{3}{c}{} & Financial  & 84 & 31 (36.9\%) & 47 (55.95\%)     \\
\multicolumn{3}{c}{} & Service  & 43 & 2 (4.65\%) & 20 (46.51\%)      \\
\multicolumn{3}{c}{} & Auxiliary & 8 & 0 (0.00\%) & 5 (62.5\%)     \\
\cmidrule(lr){4-7}
\multicolumn{3}{c}{\multirow{-7}{*}{Culpritware}} & \textbf{Total} & \textbf{250} & \textbf{71 (28.40\%)} & \textbf{149 (59.60\%)}      \\
\midrule
\multicolumn{3}{l}{Benign} & \textbf{Total}& \textbf{250} & \textbf{0 (0.00\%)} & \textbf{8 (3.20\%)}
\\
\bottomrule
\end{tabular}

\label{tab:invite}
\end{table}

%% file: body/0055_remote.tex
\subsection{The Management of Remote Servers}
\label{subsec:server}
In this section, we study the features related to the remote servers of \target{} to answer RQ3-3.
Several components are involved to manage the remote servers, including the physical servers, their IP addresses and the registered domains to facilitate users' access, which are associated with the domain registration and the domain-IP binding.
To demystify the details, we propose a framework to automatically locate and monitor possible remote servers identified from the \target{} samples.

Specifically, we first collect all URL addresses from the APK files based on known URL format (\eg, http/https format),  and filter out benign URLs using a whitelist. 
The whitelist is built by querying authoritative sources (\eg, Alexa~\cite{Alexa}) and commonly observed URLs used by the third-party libraries.
Based on the filtered URL list, we then collect \verb|DNS| and \verb|WHOIS| information for remote servers. 
After that, we connect to the remote server routinely to: 
1) crawl and store the remote resources; and
2) monitor the liveness of remote servers.
In total, we monitored $1,264$ domains every day, from January 1, 2021 to May 1, 2021.

In the following, we will first demonstrate the severity by measuring the lifespan of the remote servers of \target{} in Section~\ref{subsubsec:life-cycle}.
After that, we try to in-depth explore the strategies adopted by the operators to protect themselves in Section~\ref{subsubsec:manipulation} and Section~\ref{subsubsec:abuse}.

\begin{figure}
	\centering
	\includegraphics[width=.45\textwidth]{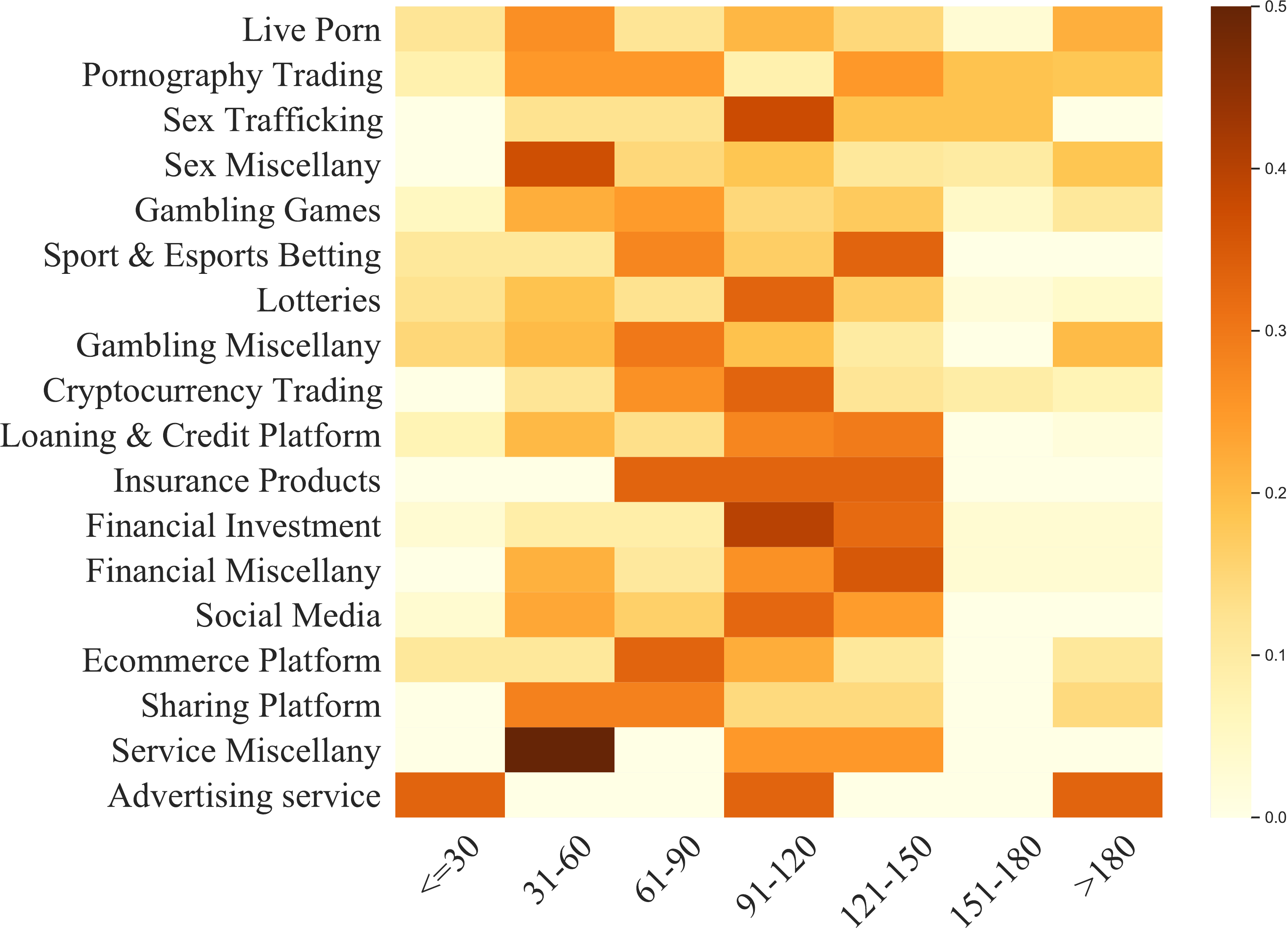}
	\caption{The lifespan of \target{}}
	\label{fig:life_cycle}
\vspace{-1em}
\end{figure}

\subsubsection{Lifespan of Remote Servers}
\label{subsubsec:life-cycle}
To evaluate the stability of the services of \target{}, we focus on the lifespan of these remote servers.
The start and end time of the lifespan is determined as follows:
\begin{itemize} [leftmargin=*]
	\item \textbf{Start time. } 
	To simplify the analysis, we take the last modification time of the \textit{AndroidManifest.xml} file in the APK file as the start time. It actually indicates the last time the APK file is packed and distributed.
	We assume the remote servers should be set up accordingly. 
	\item \textbf{End time. }
	It is determined by inspecting the status of remote servers,
	as the DNS resolution information and status codes of requests indicate the liveness of servers. 
	Specifically, we automatically monitor the \target{} samples routinely by intercepting network traffic during the startup.
	To avoid overestimation, the network traffic from third-party services (\eg, \verb|Umeng|~\cite{Umeng} and \verb|Google Service|~\cite{Googleservice}) are filtered out. 
\end{itemize}
Note that, for a given server, our way to measure the end time is conservative. 
Specifically, if the server is still alive after the end of our monitoring, we record the last inspection time as the end time. Our investigation shows that 416 ($32.9\%$) domains are still active at the end of the monitoring.
The majority of these domains fall into categories \textit{Gambling} ($36.78\%$) and \textit{Sex} ($25.48\%$).
Alternatively, if the server has expired before the first inspection, we also record the report time as the end time. 

The assessment lasts for $149$ days~\footnote{From December 6, 2020 to May 4, 2021.}, and the lifespan distribution of each category is summarized in Figure~\ref{fig:life_cycle}. 
Of course, the real lifespan of these remote servers will inevitably be longer due to the conservative end time used in our measurement.
The result shows that the lifespan of most servers fall into the range between $30$ to $150$ days, and only $8.3\%$ of them survive over $150$ days. 
The average lifespan for all categories is $110$ days.
Among them, the \textit{Financial} category has the shortest lifespan, \ie, $91$ days on average; while the \textit{Sex} category owns the longest, \ie, $125$ days on average.

Obviously, \textit{the lifespan leaves a relatively long time window for \target{} to scam victims, which indicates the failure of supervision}.
Thus there is a pressing need to tackle these remote servers to perform the protection.

\begin{figure}
    \centering
    \includegraphics[width=.4\textwidth]{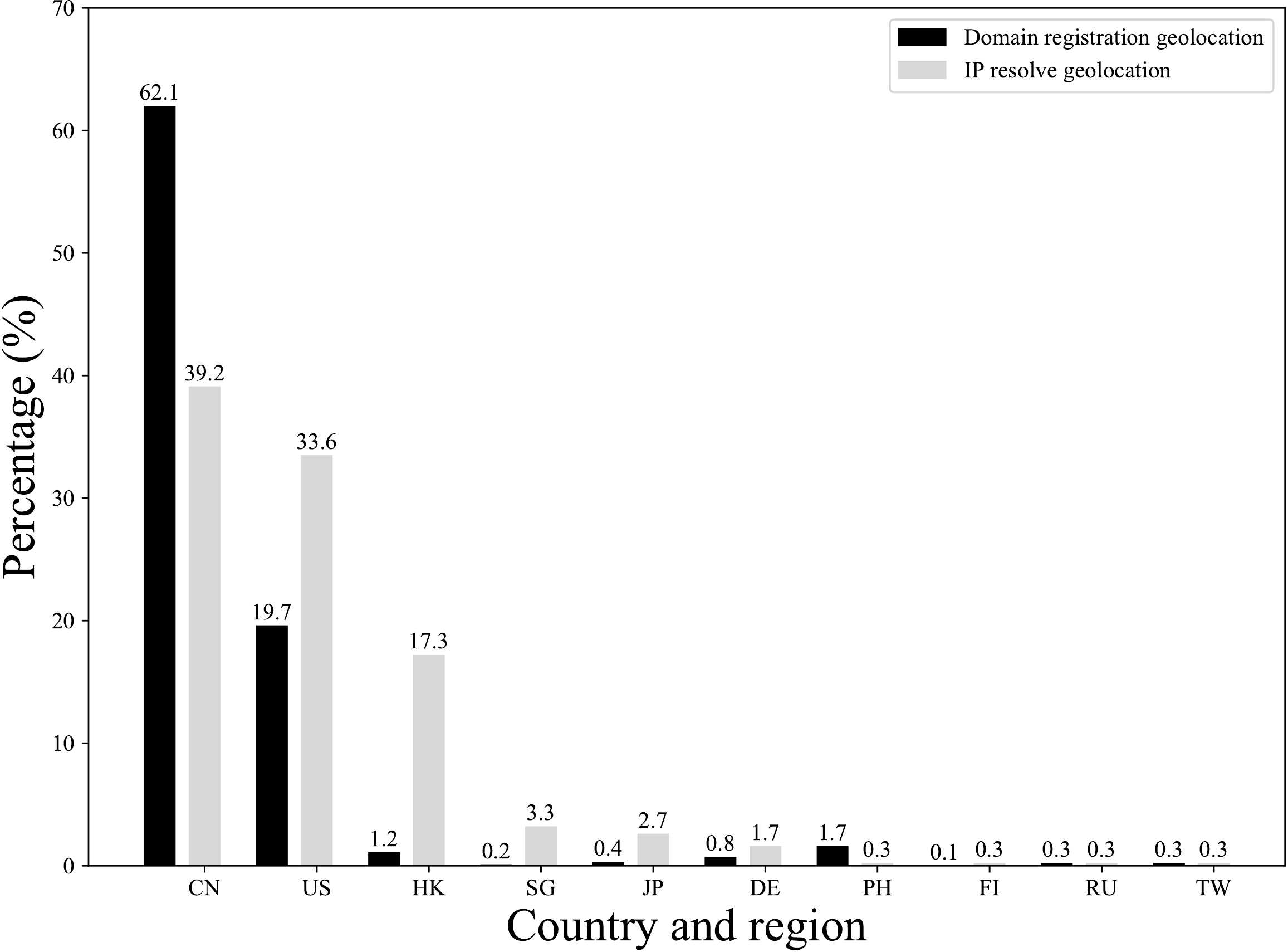}
    \caption{Top 10 countries and regions ranked by number of domain and IP geolocation.}
    \label{fig:geolocation}
\end{figure}

\subsubsection{Manipulate Domains and IP addresses}
\label{subsubsec:manipulation}
\smallskip 
\noindent \textbf{Geolocation Distribution}\tab  
The geographic location of the domain and the IP address represent the coordinates of the active area and the deployed area, respectively. 
\begin{itemize} [leftmargin=*]
	\item \textbf{The geolocation of  registered domains. } 
	We leverage the \verb|WHOIS| records of these domains to serve the need.
	In the case of incomplete \verb|WHOIS| records, the country-related top-level domain knowledge (\eg, \verb|.de| for German) can be used to improve the accuracy. 
	The aggregated country/territory of these domains are depicted as the bars in Figure~\ref{fig:geolocation}~\footnote{It's worth noting that most of the \target{} samples are collected in China, which may lead to some bias.}. 
	The result shows that the majority of domains are registered in Mainland China ($62.1\%$), the United State ($19.7\%$) and Hong Kong ($1.2\%$), respectively.
	Clearly, the majority of the \target{} samples are active in China, hence the operators are prone to register domains in China to promote the accessibility of the target users.

	\item \textbf{The geolocation of IP addresses. } 
	We first collect IP addresses of domains from DNS services. 
	Then we query the \verb|MaxMind|~\cite{maxmind} database for IP geolocation information. 
	Since the IP resolved by DNS changes frequently, we record all the IP addresses ever resolved by DNS services for a given domain.
	The bars in Figure~\ref{fig:geolocation} give the geolocation distribution.
	The result shows that the top 3 of the IP addresses' geolocation are Mainland China ($39.2\%$), the United State ($33.6\%$) and Hong Kong ($17.3\%$), respectively.
\end{itemize}

By comparing the two results in Figure~\ref{fig:geolocation}, we find that the ratio of IP addresses located in China is lower than that of the domains ($62.1\%$ vs. $39.2\%$). 
Conversely, the ratio of IP addresses located in the United State and Hong Kong are much higher.
The difference indicates that, for \target{}, there exists a deviation between the active area and the corresponding deployed area. 

\begin{figure}[!t]
	\centering
	\includegraphics[width=.4\textwidth]{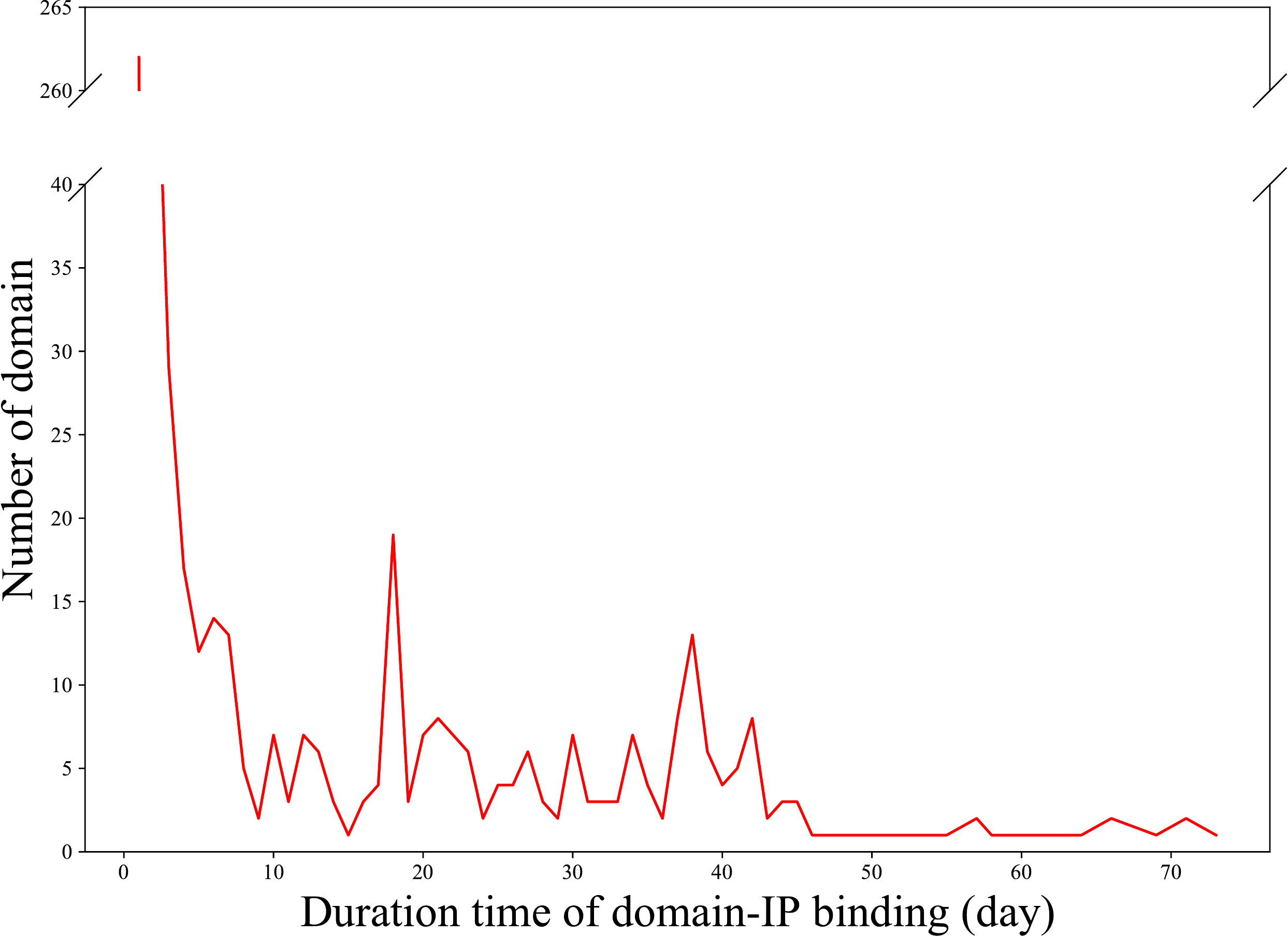}
	\caption{The domain numbers of different duration time of domain-IP binding. Note that y-axis labels between 40 and 260 are cut off due to the limited space.}
	\label{fig:ip_day}
\end{figure}

\smallskip 
\noindent \textbf{Domain-IP Relationship}.
The binding relationship between domains and IP addresses represents the mapping between the physical servers and the registered domain names.
Generally, the domain names are hard-coded in the apps, while the bound IP addresses are flexible and can be changed.
Namely, the duration time of the binding relationship of these \target{} samples may vary depending on the circumstances, as the distribution shown in Figure~\ref{fig:ip_day}.
Specifically, $307$ domains ($24.29\%$ of all) have changed the binding IP addresses at least once in two months. We call them \textit{flexible domains}.
For these domains, the average duration time of the binding is only $11.65$ days (and the shortest is even less than 1 day), which suggests that the relationship is extremely flexible.
The remaining domains ($75.71\%$ of all) are called \textit{fixed domains}, which are associated with the fixed IP addresses. Therefore their binding relationship is fixed.

Interestingly, our investigation shows that there exist two types for these flexible bindings:
1) \textbf{type-I: }multiple domains point to one IP address;
2) \textbf{type-II: }others that do not belong to type-I, which means there does not exist any special relationship among the bindings.
Among these $307$ flexible domains, $207$ (\ie, $67.43\%$) are type-I, while the remaining $100$ (\ie, $32.57\%$) are type-II. 
In particular, we observe that $43$ domains in type-I flexible bindings are to the same IP address during the same period.

\begin{figure}[!t]
	\centering
	\includegraphics[width=.5\textwidth]{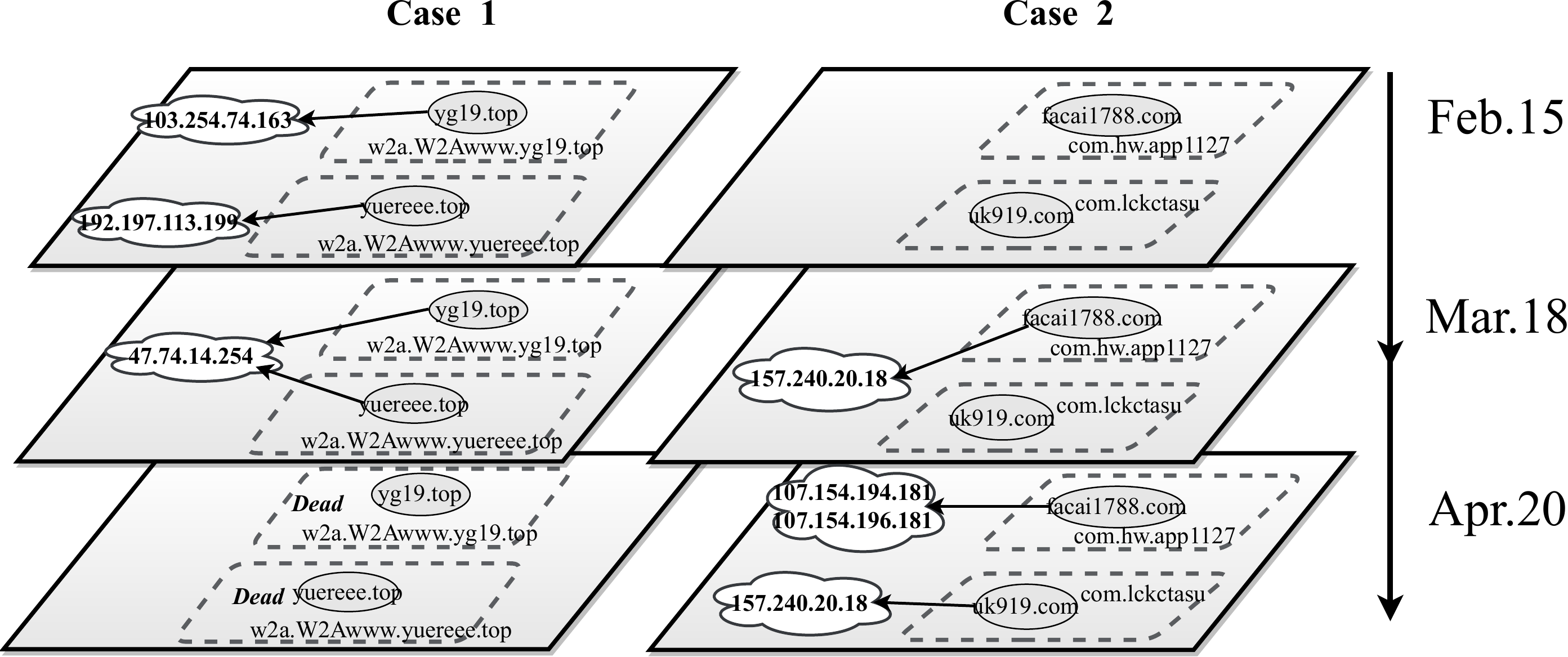}
	\caption{Examples of the flexible domain-IP bindings.}
	\label{fig:domain_relationship}
\vspace{-1em}
\end{figure}

The type-I flexible bindings are relatively complicated, and can be divided into two cases.
To provide a better illustration, we present concrete examples in Figure~\ref{fig:domain_relationship}.
Specifically, the left part of Figure~\ref{fig:domain_relationship} gives an example of the first case, \ie, multiple domains point to one IP address \textit{during the same time period}.
The domain \code{yg19.top} (used in \code{w2a.W2Awww.yg19.top}) and \code{yuereee.top} (used in \code{w2a.W2Awww.yuereee.top}) associate to the IP address \code{47.74.14.254} on March 18, 2021.  
While the right part of Figure~\ref{fig:domain_relationship} gives an example of the second case, \ie, multiple domains point to one IP address \textit{in different time periods}. 
The domain \code{uk919.com} (used in \code{com.lckctasu}) associates to IP address \code{157.240.20.18} on April 20, 2021. However, this IP address has been associated with \code{facai1788.com} (used in \code{com.hw.app1127}) on March 18, 2021.

Due to the conservative essence, the lifespan calculated in Section~\ref{subsubsec:life-cycle} is not appropriate to evaluate the advantage of the flexible bindings over those fixed ones.
However, we observe that domains associated with the flexible bindings are more likely to be alive after the monitoring to calculate the end time of the lifespan.
As a matter of fact, for all domains, the flexible ones only occupy $24.29\%$; while for the active domains after the monitoring, the flexible ones occupy $48.08\%$.
The result suggests that \textit{the tangled and mutable bindings may complicate the regulatory authorities' efforts to trace them}.

\subsubsection{Abuse Domain Registrants}
\label{subsubsec:abuse}
We rely on \verb|WHOIS| records to analyze the registrants. 
In total, we collect $164$ different registrants from $1, 264$ domains.
We find that most domain registrants of \target{} are using \verb|GoDaddy| and \verb|Alibaba Cloud|.
The registrants that contain more than $15$ domains are listed in Table~\ref{tab:Registrar}.
The total number of domains in the table is $896$, occupying $70.89\%$ of all domains.
The result demonstrates that these domain service providers have been abused.

\input{tables/domain_proxy}

\begin{tcolorbox}[size=title]
\textbf{Finding \#6:}
The average lifespan of those remote servers is $110$ days, indicating a long time window to scam victims. 
The strategies adopted by the operators include actively manipulating domains and IP addresses, and abusing domain registrants.
More effective supervision is required to tackle \target{}. 
\end{tcolorbox}

%% file: tables/domain_proxy.tex
\begin{table}[!t]
\centering
\footnotesize
\setlength{\abovecaptionskip}{-1ex}
\caption{The registrants of domains.}
\begin{tabularx}{.4\textwidth}{lXl} 
\toprule
 \textbf{Registrant} & \textbf{Count} & \textbf{Percentage} \\
\midrule  
Alibaba Cloud (Beijing)  & 279 & 22.07\%  \\ 
GoDaddy.com, LLC &  272 & 21.52\% \\ 
Xin Net Technology Corporation & 68 & 5.38\% \\
Alibaba Cloud (WanWang)  & 51 & 4.03\% \\
NAMECHEAP INC & 37 & 2.93\% \\
NameSilo, LLC & 36 & 2.85\% \\
eName Technology Co.,Ltd. & 34 & 2.69\% \\
Name.com, Inc & 30 & 2.37\% \\
MarkMonitor, Inc. & 21 & 1.67\% \\
Network Solutions, LLC & 19 & 1.50\% \\
GANDI SAS & 17 & 1.34\% \\
Amazon Registrar, Inc. & 16 &1.27\% \\
Chengdu west dimension dt& 16 & 1.27\% \\
\midrule
 \textbf{Total} & \textbf{896} & \textbf{70.89\%} \\
\bottomrule 
\end{tabularx}
\setlength{\belowcaptionskip}{0ex}
 \label{tab:Registrar}
\end{table}

%% file: body/0056_payment.tex
\subsection{Payment Analysis} 
\label{subsec:paymentservices}
In this section, we characterize the payment services of \target{} to answer RQ3-4.
To this end, we randomly select 25 \target{} samples from the \abnormal{} dataset, and manually investigate the payment process by first initiating payment requests and then analyzing them.

Specifically, the bank transaction and digital-currency is easy to figure out. To distinguish the third-party payment and fourth-party payment, we first build a list of domains used by licensed third-party payment services.
For any domain in this list, it only points to a single merchant as a normal third-party payment service.
Otherwise, we consider it as a candidate for a fourth-party payment.
Then we make multiple payment requests with different amounts to determine the payment type. 
If the recipient accounts are changed frequently, we consider it as a fourth-party payment. 

By using the above approach,  we identify fourth-party payment services and third-party payment services across 25 \target{} samples,
which are listed in Table~\ref{tab:payment} in Appendix.
We observe that:
1) one sample merely uses the third-party payment;
2) two samples support both payment methods, and randomly choose one of them; and
3) 22 samples only rely on the fourth-party payment.
The result shows that \textit{most \target{} ($96\%$) adopt fourth-party payment services.
}

\begin{figure}[!t]
\centering
    \subfigure[]{
    \begin{minipage}[t]{1in}
        \label{fig:none-sdk-3}
        \includegraphics[width=1in]{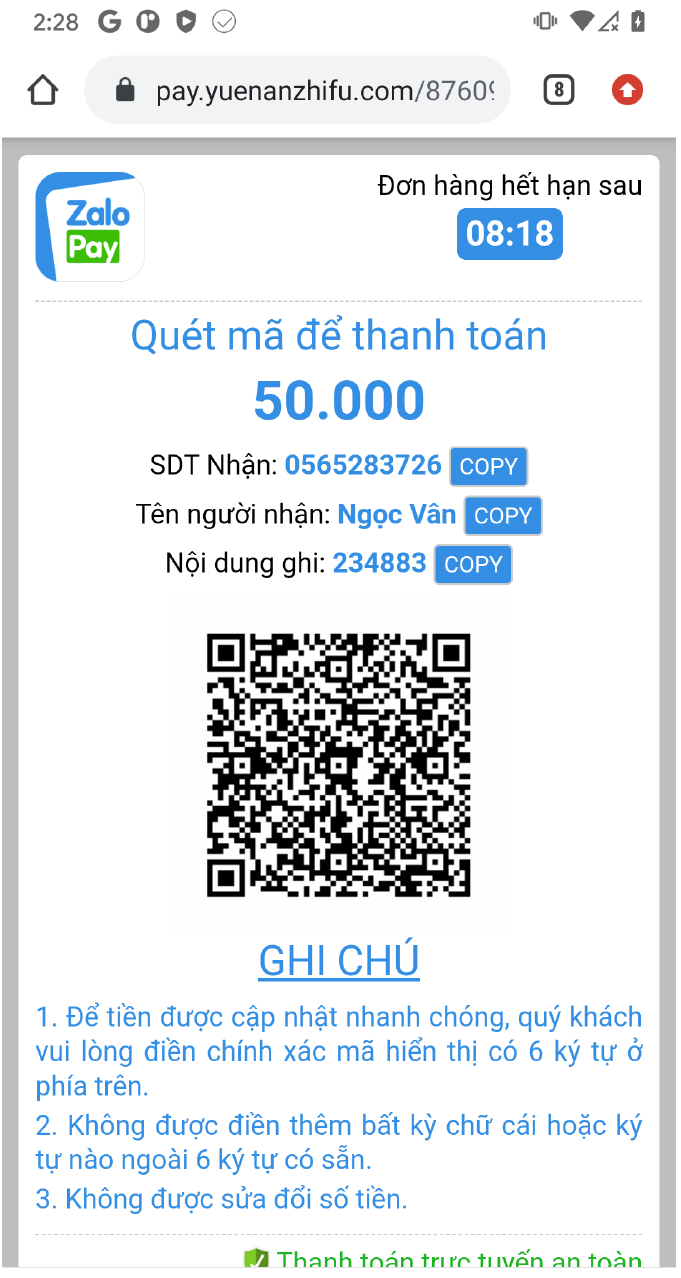}
    \end{minipage}}
    \subfigure[]{
    \begin{minipage}[t]{1in}
       \label{fig:none-sdk-2}
       \includegraphics[width=1in]{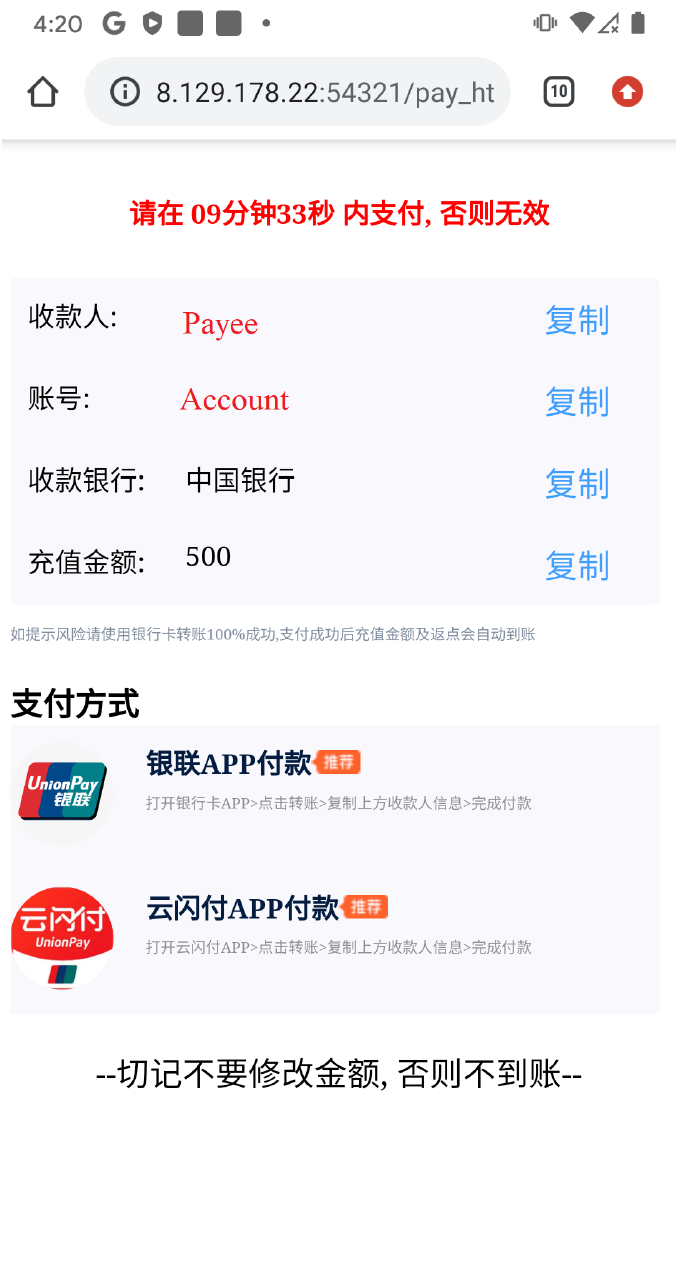}
    \end{minipage}}
    \subfigure[]{
    \begin{minipage}[t]{1in}	
        \label{fig:none-sdk-1}
        \includegraphics[width=1in]{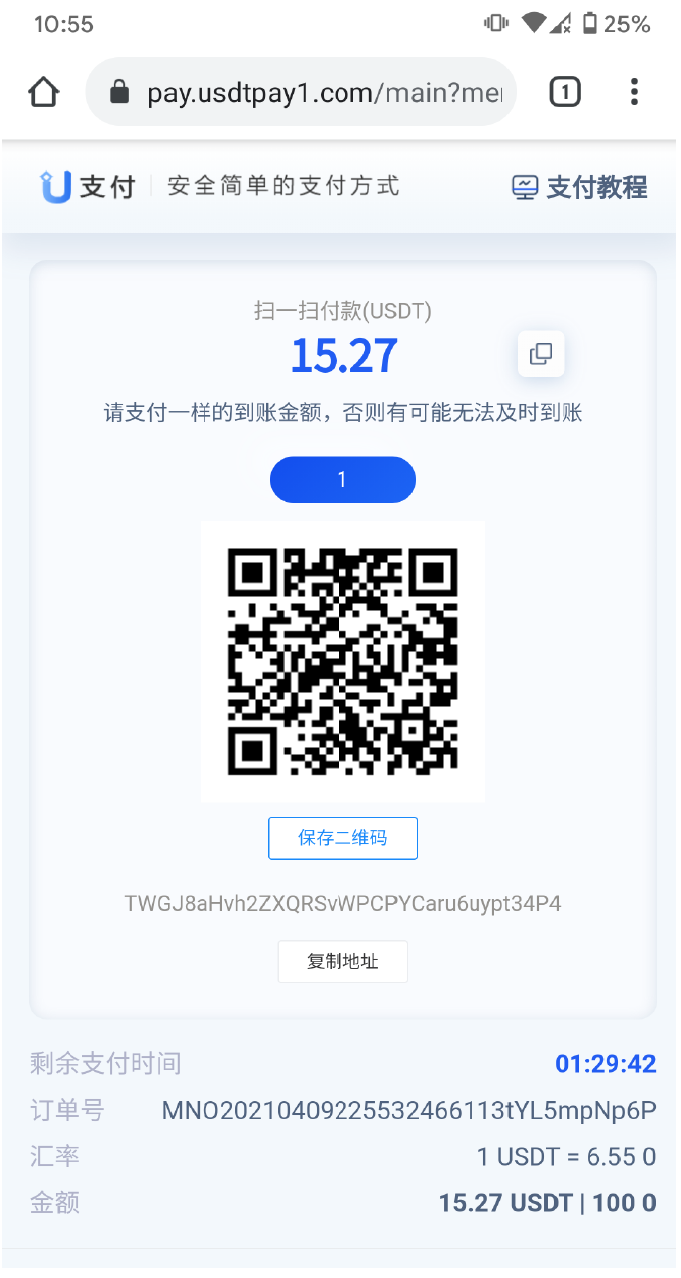}
    \end{minipage}}
 \caption{ Fourth-party payment services based on different formal payment.(a) based on third-party payment. (b) based on bank transactions. (c) based on digital-currency.}
 \label{fig:none-sdk}
\vspace{-1em}
\end{figure}

Note that the fourth-party payment may use different types of payments, \ie, the \textit{third-party} payment, the \textit{digital-currency} payment, and the traditional \textit{bank transaction} payment.
Examples of fourth-party payments are depicted as Figure~\ref{fig:none-sdk}. 
The sub-figures, from left to right, represent the fourth-party payment based on the third-party payment, the bank transaction payment and the digital-currency payment, respectively.
To further demystify the fourth-party payment, we analyze the payments used in the fourth-party payment services from these 25 samples.
In total, we identify $47$ different fourth-party payment services, which rely on the payment services listed in  Table~\ref{tab:fourth_party_pay} in Appendix. The result shows that most of the fourth-party payment services ($65.96\%$) are based on third-party payment. While the bank transaction payments occupy $23.40\%$ and the digital-currency ones occupy $8.51\%$.

Besides, more interestingly, we observe that \textit{the fourth-party payment does not use the formal payment directly} (\eg, invoking the APIs provided by the third-party payment). Instead, it requires the user to do it using the following two steps, \ie, first recording the transaction information (\eg, the transfer destination), and then finalizing the payment by feeding the formal payment with the recorded information. 
Such a behavior is probably due to the strategy adopted by \target{} to hide their traces.

\begin{tcolorbox}[size=title]
\textbf{Finding \#7:}
Most \target{} ($96\%$) adopt covert fourth-party payment services, which indirectly use the third-party payments by requiring the users to finalize the payments. Specifically, these fourth-party payments rely on the third-party payments ($65.96\%$), the bank transactions ($23.40\%$) and the digital-currency payments ($8.51\%$), respectively. 
\end{tcolorbox}

%% file: body/006_discussion.tex
\section{Discussion}
Due to the covert distribution channels, it is quite difficult to obtain the \target{} samples directly. 
Instead, we rely on the authority department to collect samples. They are provided by victims. This is a slow and incremental process.
As a result, the number of the collected samples is limited. 
Based on the understanding of this work, it is possible to collect samples by monitoring the distribution channels discussed in Section~\ref{subsec:propagation} directly in the future.

Some analyses conducted in this paper are based on manual efforts, and the scalability is inevitably limited.
For example, we only manually investigate top 10 app groups for the provenance of the developers in Section~\ref{subsubsec:provenance}. This issue can be solved by building a tool which is capable of automatically crawling and analyzing data from the online sources.

Besides, our study does not cover in-depth code features of apps. 
Although we find that the developers are more inclined to develop hybrid apps, and they may deploy code with sensitive functionalities in the remote servers.
However, there may still exist sensitive code in apps, \eg, the third-party (native) libraries used by \target{}, which require more effort to delve into the details.

%% file: body/007_relatedwork.tex
\section{Related Work}
\subsection{\Ccrime{} Analysis}
Maciej et al.~\cite{korczynski2018cybercrime} presented the study of abuse in the domains registered and pointed out that cyber-criminals increasingly prefer to register, rather than hack domain names.
Hu et al.~\cite{hu2018dating} analyzed the ecosystem of romantic dating apps and revealed the purpose, \ie, luring users to get profits. 
Roundy et al.~\cite{roundy2020many} focused on the Creepware apps that launch interpersonal attacks.
Stajano et al.~\cite{stajano2011understanding} examined various scams and extracted the general principles of scam strategies.
Blaszczynski et al.~\cite{banks2016online} characterized the relationship between gambling and crime, and concluded the behavior of illegal gambling.
To the best of our knowledge, none of the previous works systematically study \target{} and the ecosystem.

\subsection{Security Analysis for Android Apps}
\textbf{Malware detection. } 
Sun et al.~\cite{sun2016sigpid} studied the permission requirements of Android apps and proved that this can be used to distinguish malware.
Arora et al.~\cite{arora2019permpair} proposed a detection model by extracting the permission pairs from the manifest file.
Chen et al.~\cite{chen2017uncovering} characterized existing Android ransomware and proposed a real-time detection system by monitoring the user interface widgets and finger movements.

\noindent
\textbf{Vulnerability detection. }
Chan et al.~\cite{chan2011privilege} proposed a vulnerability checking system to check vulnerabilities may be leveraged by privilege escalation attack.
Joshi et al.~\cite{joshi2016android} systematically elaborated the vulnerabilities that exist in the Android operating system.

\noindent
\textbf{Similarity detection.}
Sebastian et al.~\cite{sebastian2020towards} presented an approach based on ownership for identifying developer accounts in mobile markets.
Chen et al.~\cite{chen2014achieving} presented graphical user interface similarity to detect application clones.

%% file: body/008_conclusion.tex
\section{Conclusion}
In this paper, we take the first step to systematically demystify \target{} and the underlying ecosystem.
Specifically, we first characterize \target{} by conducting a comparison study with \normal{} apps and \malware{}.
Then we reveal the structure of the ecosystem, including the participating entities and the workflow.
Finally, we perform an in-depth investigation to demystify the characteristics of the participating entities of the ecosystem.
Our research reveals some impressive findings which may help the community and law enforcement authorities protect against the emerging threats.

%% file: body/appendix.tex
\section{Appendix}
\label{appendix}

\subsection*{Profit Tactics}
\label{appendix:profit_tricks}
Here we summarize the 12 profit tactics, as follows: 
\begin{enumerate} [leftmargin=*]
    \item \textbf{Misrepresentation:} By leveraging false information, the apps convert the victims to trust they are formal and get profit.
    \item \textbf{Romance Fraud:}  Apps provide live-streaming rooms or chatting rooms for perpetrators to induce money from the victims by pretending to keep a romantic relationship with the victims.
    \item \textbf{Black-Box Operation:} Apps design many games and lure victims into playing. But due to the imbalance of the game, the victims will lose all cash.
    \item \textbf{Low quality Delivery:} Apps falsify apocryphal products that do not exist or describe products with fake statements that don't match the fact. Victims make the payment but the quality or quantity of the goods or services received is much less than described.
    \item \textbf{Advanced Fee:} Apps notify victims that they are eligible for a large financial fund, but must first pay a tax or fee to get the fund. The victims paid the advance but never not receive the money promised.
    \item \textbf{Investment Scam:} Apps seduce investors into buying investment products by offering false information about them that offer huge returns with little risk. 
    \item \textbf{Credit Card Fraud:} Victims used a credit card to pay for a transaction on apps but found that their credit card had been stolen.
    \item \textbf{Propagation of Misinformation:} Apps take money from certain organizations or individuals to spread harmful information (violent speech, fake news, etc.)
    \item \textbf{Accessory Act:} These apps are not engaged in any criminal act directly, but promote other illegal behaviors and get a commission from their revenue.
    \item \textbf{Extortion:} Victims received intimidation or improper exercise of power after using apps with the intent of illegally extracting money or property. It may include threats of physical harm, criminal prosecution, or public exposure.
    \item \textbf{Personal Data Breach:} Apps collect the personal data of victims and sell it to get benefits.
\end{enumerate}

\input{tables/behaviour}

\input{tables/clustering_process}

\input{tables/operator}

\input{tables/payment}

\input{tables/Fourth_party}

\begin{table*}[h]
\centering
\begin{threeparttable}
\caption{Summary of app generators in the ecosystem (see Section~\ref{subsubsec:observation}).}
\label{tab:App_generator_collection}
\centering
\footnotesize
\begin{tabular}{llcccc}
\toprule
\hline
\textbf{App Generator} & Official Website & Local Resources & Remote Services  & Prerequisites & Encryption \\ 
\midrule
\textbf{DCloud} & \url{https://dcloud.io/} & $\bullet$ & - & - & - \\ 
\textbf{APICloud} & \url{https://www.apicloud.com/} & $\bullet$ & - & -& RC4 (Java) \\
\textbf{BSLApp} &\url{https://www.appbsl.cn/} & - & $\bullet$ & -  & AES.CBC (Java) \\
\textbf{Cordova} &\url{https://cordova.apache.org/} &$\bullet$ & - &-  & - \\
\textbf{Ionic} &\url{https://ionicframework.com/} &$\bullet$ & - & AngularJS & - \\
\textbf{Onsen UI} &\url{https://onsen.io/} &$\bullet$ & - & AngularJS & - \\
\textbf{Framework 7} &\url{https://framework7.io/} &$\bullet$ & - & HTML, JavaScript & - \\
\textbf{React Native} &\url{http://www.reactnative.com/} &- & - & React & - \\
\textbf{jQuery Mobile} &\url{https://jquerymobile.com/} &- & - & HTML, jQuery & - \\
\textbf{Native Script} &\url{https://www.nativescript.org/} &$\bullet$ & $\bullet$ & JavaScript & - \\
\textbf{Famous} &\url{https://famous.co/} &$\bullet$ & - & WebGL, AngularJS & - \\
\textbf{Intel XDK} &\url{https://software.intel.com/} &- & - & - & - \\
\textbf{Sencha Touch} &\url{http://www.sencha.com/} &$\bullet$ & - &-  & - \\
\textbf{Kendo UI} &\url{http://www.telerik.com/} &$\bullet$ & - &-  & - \\
\textbf{Mobile Angular} &\url{http://mobileangularui.com/} &$\bullet$ & - &-  & - \\
\textbf{Monaca} &\url{https://monaca.io/} &$\bullet$ & - &-  & - \\
\textbf{Trigger.IO} &\url{https://trigger.io/} &$\bullet$ & $\bullet$ &-  & - \\
\textbf{Seattle Cloud} &\url{http://seattleclouds.com/} &$\bullet$ & $ -$ &- & -\\
\textbf{Andromo} &\url{http://www.andromo.com/} &$\bullet$ & $\bullet$ &- & -\\
\textbf{Apps Geyser} &\url{http://www.appsgeyser.com/} &$\bullet$ & $\bullet$ &- & -\\
\textbf{Biznessapps} &\url{http://www.biznessapps.com/} &- & $\bullet$ &- & -\\
\textbf{Appinventor} &\url{http://appinventor.mit.edu/} &$\bullet$ & $\bullet$ &- & -\\
\textbf{AppYet} &\url{http://www.appyet.com/} &$\bullet$ & $\bullet$ &- & DES.CBC (Java)\\
\textbf{Mobincube} &\url{http://www.mobincube.com/} &$\bullet$ & $\bullet$ &- & -\\
\textbf{Appy Pie} &\url{http://www.appypie.com/} &$\bullet$ & - &-  & - \\
\textbf{Appmachine} &\url{http://www.appmachine.com/} &$\bullet$ & - &-  & C\# (TEA) \\
\textbf{Good Barber	} &\url{http://www.goodbarber.com/} &$\bullet$ & $\bullet$ &-  & - \\
\textbf{Shoutem} &\url{http://www.shoutem.com/} &- & $\bullet$ &-  & - \\
\textbf{Apps Builder} &\url{http://www.apps-builder.com/} &$\bullet$ & - &-  & - \\
\textbf{Appmakr} &\url{http://appmakr.com/} &$\bullet$ & - &-  & - \\
\textbf{appery.io} &\url{https://appery.io/} &$\bullet$ & - &-  & - \\
\textbf{Xamarin} &\url{https://docs.microsoft.com/} &- &$-$ &.NET  & - \\
\textbf{PhoneGap} &\url{https://phonegap.com/} &- &- &HTML, JavaScript  & - \\
\textbf{bufanapp} &\url{https://www.bufanapp.com/} &- &$\bullet$ &-  & RC4 (Java) \\
\textbf{AppCan} &\url{http://www.appcan.cn/} &$\bullet$ &$\bullet$ &-  & RC4 (Native) \\
\textbf{Dibaqu} &\url{https://www.dibaqu.com/} &- &- &-  & RC4 (Native) \\
\textbf{Pgyer} &\url{https://www.pgyer.com/} &- &- &-  & RC4 (Native) \\
\textbf{ofcms} &\url{https://www.ofcms.com/} &- &$\bullet$ &-  & - \\
\textbf{ChuXueYun} &\url{https://www.chuxueyun.com/} &- &$\bullet$ &-  & AES.CBC (Java) \\
\textbf{AppPark} &\url{http://www.apppark.cn/} &$\bullet$ &$\bullet$ &-  & - \\
\textbf{SuishouApp} &\url{http://www.suishouapp.com/} &- &$\bullet$ &-  & AES.CBC (Java)\\
\textbf{Hema} &\url{https://hema.im/} &- &- &-  & -\\
\textbf{Justep} &\url{https://www.justep.com/} &$\bullet$ & - &-  & - \\
\textbf{YunDaBao} &\url{http://app.yundabao.cn} &$\bullet$ & - &-  & - \\
\textbf{yunedit} &\url{https://www.justep.com/} &$\bullet$ & - &-  & - \\
\textbf{apkeditor} &\url{https://apkeditor.cn/} &$\bullet$ & $\bullet$ &-  & - \\
\textbf{yimen} &\url{https://www.yimenapp.net/} &$\bullet$ & $\bullet$ &-  & AES.CBC (Java) \\
\hline
\bottomrule
\end{tabular}
\begin{tablenotes}
\item[1] {For the developer's code encrypted by the app generator, we either use the decryption process in app (Java code) or simulate~\cite{AndroidNativeEmu} the decryption process in app (native code) to get the original developer's code.}.

\end{tablenotes}

\end{threeparttable}
\end{table*}

%% file: tables/behaviour.tex
\begin{table*}[h]
\footnotesize
    \centering
    \caption{Categorization criteria of \target{}.}
\label{tab:App_category}
\begin{threeparttable}
\begin{tabular}{llccccccccccl}
\toprule
& & & \multicolumn{3}{c}{User Seducement}& \multicolumn{3}{c}{Purchase \& Deposit} & \multicolumn{3}{c}{Follow-up }
\cr\cmidrule(lr){4-6}\cmidrule(lr){7-9}\cmidrule(lr){10-12}
Top-category & Sub-category & Portion & U1 & U2 & U3 & D1 & D2 & D3 & F1 & F2 & F3 & Profit Tactic\tnote{1} \cr  
\midrule
\multirow{4}{*}{\textbf{Sex}} &
\textbf{Live Porn} & 4.98\% &  &  & \faCircle & \faCircle &  &  &  &  & \faCircle & P2, P10, P11 \\
& \textbf{Pornography Trading} & 1.78\% & \faCircle  &  & \faCircleO &  & \faCircle &  & \faCircle &  &  & P4 \\ 
& \textbf{Sex Trafficking} & 2.25\% & \faCircle &  &  &  & \faCircleO &  &  \faCircle & &  \faCircle &  P2 \\ 
& \textbf{Sex Miscellany} & 4.03\% & & & & & & & & & & \\ \hline
\multirow{4}{*}{\textbf{Gambling}} &
\textbf{Gambling Games} & 17.91\% &  &  & \faCircle &  &  & \faCircle &  & \faCircle &  & P3, P11 \\ 
& \textbf{Sports \& E-sports Betting} &  4.03\% & \faCircle  &  &  &  & &\faCircle  &  & \faCircle &  & P3, P11 \\ 
& \textbf{Lotteries} &  8.07\% & \faCircle  &  &  & & \faCircle &  & \faCircle & \faCircleO &  & P1, P3 \\ 
& \textbf{Gambling Miscellany} &  0.95\% & & & & & & & & & & \\\hline
\multirow{5}{*}{\textbf{Financial}} &
\textbf{Cryptocurrency Trading} & 6.41\% & \faCircle  &   & & \faCircleO   & \faCircle & &   &  \faCircle &  & P6 \\ 
& \textbf{Loan \& Credit Platform} & 17.32\% & \faCircle  & \faCircle  & &   & \faCircle   & &  \faCircle  & &  & P1, P5, P11 \\ 
& \textbf{Insurance Products} & 0.36\% & \faCircle  & \faCircle  & & & \faCircle  &  & \faCircle  & &  & P1, P4, P7 ,P11 \\ 
& \textbf{Financial Investment} & 9.25\% & \faCircle   & \faCircle  & & & \faCircle  &  &  & \faCircle  &  & P1, P6, P9 \\ 
& \textbf{Financial Miscellany} & 8.90\% & & & & & & & & & & \\ \hline
\multirow{4}{*}{\textbf{Service}} &
\textbf{Social Media} &  8.90\% &  & \faCircle & &  &  & & &  & \faCircle  & P1, P2, P8, P11 \\ 
& \textbf{Ecommerce Platform} & 0.83\% & \faCircle   &\faCircleO & &  & \faCircle  &  & \faCircle  & & & P1, P4 \\ 
&\textbf{Sharing Platform} &  1.07\% & \faCircle   &\faCircleO  & &\faCircleO  &   &  &  & & & P8 \\ 
& \textbf{Service Miscellany} & 2.02\% & & & & & & & & & & \\ \hline
\textbf{Auxiliary Tool} &
\textbf{Advertising service} &  0.95\% &  & \faCircle & &  &   &  &  & & & P1, P9 \\ 
\bottomrule
\end{tabular}
\begin{tablenotes}
\footnotesize
\item[1] The P1-P11 represent the 11 different profit \trick{}s which are detailed in Appendix. 
\item[2] Explanation of different behaviour phases:\\
\textbf{User Seducement Behaviour}: \\
\textit{U1: Fake information.} whether to leverage false or exaggerated product information  or customer information. \\
\textit{U2: Disguised.} whether disguised as regular applications or formal companies' products. \\
\textit{U3: Free Trial.} whether offer the opportunity to try it for free in a limited period.\\
\textbf{Purchase \& Deposit Behaviour}: \\
\textit{D1: Activation fee.} whether to require users to pay an activation fee first before using. \\
\textit{D2: Product/Service.} whether offer a product or service for users to purchase. \\
\textit{D3: Token purchase.} whether own currencies specifically designed for their functionality.\\
\textbf{Follow-up Behaviour}: \\
\textit{F1: None or low-quality delivery.} whether the goods or services received were of a measurably lesser quality or quantity than described. \\
\textit{F2: Disable functionality.} whether block users from logging in or withdrawing money. \\
\textit{F3: Contact interrupted.} whether to interrupt the contact with the user.\\
The two symbols in the table show whether the given behaviour is of major (\faCircle) or minor (\faCircleO) importance for the given app category.
\end{tablenotes}
\end{threeparttable}
\end{table*}

%% file: tables/clustering_process.tex
\begin{algorithm}[htb]
    \footnotesize
	\SetAlgoLined
	\SetKwData{True}{true}
  
	\KwData{$samples$: the collected \target{} samples.}
	\KwResult{$G=(Nodes,Edges,Weights)$: the generated association graph for $samples$.}
	
    \SetKwFunction{FAssociate}{associate}
    \SetKwFunction{FPreprocess}{preprocess}
    \SetKwFunction{FAssociateSamples}{associate\_samples}
    \SetKwFunction{FGetSignature}{GetSignature}
    \SetKwFunction{FGetURLList}{GetURLList}
    \SetKwFunction{FGetSnapshot}{GetSnapshot}
    \SetKwFunction{FAssociateSignature}{AssocSignature}
    \SetKwFunction{FAssociateURLList}{AssocURLList}
    \SetKwFunction{FAssociateSnapshot}{AssocSnapshot}
     \SetKwFunction{Fdoassociate}{do\_associate}
    \SetKwProg{Fn}{Function}{:}{}
    
    \Fn{\FAssociateSamples{$samples$}} {
        $features \gets \FPreprocess()$\;
        \tcp{feature[i] = (sig, url\_list, snapshot)}
        \For{$sample_i$ in $samples$} {
            $I$ = 0; $N$ = $N'$ = $\{sample_i\}$\; 
            
            \While{$I \leq I_{max}$} {
                \For{$n$ in $N'$} {
                    $N'$ = \FAssociate($n$, $samples$, $features$[$n$])\; 
                    $N$ = $N~\cup$~$N'$\;
                }
                $I += 1$
            }
            
            \For{$n$ in $N$} {
                $edges$~= $edges~\cup~(sample_i, n)$\;
            }
            $nodes$~= $nodes~\cup~sample_i$\;
        }
        \KwRet $G(nodes, edges)$\;
    }
    
    \Fn{\FAssociate{$seed$, $samples$, $feature$}} {
        $assoc$ = $\{\}$\;
        $assoc$ += \Fdoassociate~($samples$, $feature$.$sig$)\tcp*{~\ding{182}}
        $assoc$ += \Fdoassociate~($samples$, $feature$.$url\_list$)\tcp*{~\ding{183}}
        $assoc$ += \Fdoassociate~($samples$, $feature$.$snapshot$)\tcp*{~\ding{184}}
        \KwRet $assoc$\;
    }
    
    
	\caption{The association algorithm}
  \label{alg:cluster}
\end{algorithm}

%% file: tables/operator.tex
\begin{table*}[h]
\centering
\small
\caption{The operators in the interaction phase.}
\begin{threeparttable}
\begin{tabularx}{0.95\textwidth}{lll}
\toprule 
 \textbf{Operator} & \textbf{Description } & \textbf{Profit Tactic \tnote{1} }\\
\midrule  
Porn Seller  & Communicates with victims and uses romantic traps to make revenue.  & P2\\ 
Scammer &  Uses verbal tricks to persuade the victims to do harmful behaviors. & P6\\ 
Rumormonger  & Be hired by certain organizations to public false information that confuses victims. & P8 \\
Blackmailer & Threatens the victims after getting the victims' sensitive personal information. & P10  \\
\bottomrule 
\end{tabularx}
\begin{tablenotes}
\footnotesize
\item[1] The profit \trick{} where operators are active in. The details of profit \trick{}s are listed in Appendix.
\end{tablenotes}
\end{threeparttable}
\label{tab:operators}
\end{table*}

%% file: tables/payment.tex
\begin{table}[h]\scriptsize
\centering

\caption{Payment service analysis.}
\label{tab:payment}
\begin{tabularx}{.4\textwidth}{lcccc}
\toprule 
 \textbf{Package Name} & \textbf{FP}  & \textbf{TP} & \textbf{BT} & \textbf{DC} \\
\hline  
com.jijiehaoqipai.jjhqp & $\bullet$  & -  & -  & -  \\
com.ibxcisekdcas.bieidhs & $\bullet$  & -  & -  & - \\
com.game.qebjsa.and & $\bullet$  & - & -  & - \\
com.hwgjagwgaw.hwhga & $\bullet$  & - & -  & - \\
com.game.qebr.and & $\bullet$  &  - & -  & - \\
com.df.bwtnative.op3052  & $\bullet$ & -  & -  & - \\
com.df.bwtnative.op1014  & $\bullet$ & - & -  & - \\
com.qqy.oygj & -  & $\bullet$ & -  & - \\
com.dawoo.gamebox.sid1338  & $\bullet$ & - & -  & - \\
com.adult.zero  & $\bullet$ & -  & -  & - \\
com.tianbo.tbsport  & $\bullet$ & -   & -  & - \\
com.key0.build  & $\bullet$ & $\bullet$ & -  & - \\
com.android.driver.three651a1  & $\bullet$ & - & -  & - \\
com.wj2588.esport  & $\bullet$ & -  & -  & - \\
com.youzu.android.snbx  & $\bullet$ &  - & -  & - \\
meishi.ap  & $\bullet$ & - & -  & - \\
com.binli.qipai.gp  & $\bullet$ & -  & -  & - \\
com.tea.pincha51 & $\bullet$  & - & -  & -  \\
com.ch.myframe  & $\bullet$  & -  & -  & - \\
com.CCq3.Q3omed  & $\bullet$  & - & -  & - \\
plus.H5F29087B  & $\bullet$  & - & -  & - \\
com.didi.live.spring & $\bullet$   &  $\bullet$  & -  & - \\
com.master58782sds0326  & $\bullet$  & - & -  & - \\
com.fatiao.fft & $\bullet$  & - & -  & - \\
com.sagadsg.user.mada117857  & $\bullet$  & -  & -  & - \\
\bottomrule 
\end{tabularx}
\begin{tablenotes}
\footnotesize
 \item[1] FP: Fourth-party, TP: Third-party, BT: Bank Transaction, DC: Digital-currency
\end{tablenotes}
\end{table}

%% file: tables/Fourth_party.tex
\begin{table}[h]
\scriptsize
\caption{Domains of fourth-party payment.}
\begin{tabular}{lcccl}
\toprule
\textbf{Domain}&\textbf{ Third-party}& \textbf{Bank Transaction} & \textbf{Digital-currency}  \\
\midrule
\textbf{huangzi1.xyz}  & $\bullet$ &   &   \\
\textbf{qkdt.hebeiqi.com}  &   &   & $\bullet$     \\
\textbf{p.donedonehub.xyz}  & $\bullet$  &  &   \\
\textbf{api.sepay100.com}  & $\bullet$  &  &  \\
\textbf{go04.hbtongpay.com}  & $\bullet$  &   &   \\
\textbf{hxx20p6s.qcx.ink}   & $\bullet$ & &  \\
\textbf{buy500.tuppods55.com}   &  & $\bullet$  &   \\
\textbf{fpay633.5ga.xyz} & $\bullet$  &  &  \\
\textbf{cash.cnyffzdc.com}  & $\bullet$ &  &    \\
\textbf{170.33.9.23:28090} & $\bullet$   & &  \\
\textbf{114.55.109.252:8080}  & $\bullet$  &   &  \\
\textbf{47.119.117.117} & $\bullet$ & &   \\
\textbf{user.slypay.net:9001} & $\bullet$  &  &  \\
\textbf{45.113.202.20:8180} & $\bullet$  &  &   \\
\textbf{shop.hhdpyy33.com} &   & $\bullet$  &   \\
\textbf{47.113.217.241} & $\bullet$  & &  \\
\textbf{124.71.42.29:84}  & $\bullet$  &  &   \\
\textbf{shop.jxp2233.com}   &  & $\bullet$  &    \\
\textbf{shop.dali5566.com} &  & $\bullet$ &   \\
\textbf{45.113.202.124:8180}  & $\bullet$   &  &        \\
\textbf{n.wfjjcwl.com}  & $\bullet$ &   &   \\
\textbf{cprsc.com} & & $\bullet$ &  \\
\textbf{alhfp.sxwttech.com}  &  & $\bullet$  &  \\
\textbf{cyy7ju.mango100.xyz:2053}   & $\bullet$ &  &   \\
\textbf{124.71.42.29:84}  &  & $\bullet$  &  \\
\textbf{aaa.yataiqc.com} &  &  &   \\
\textbf{120.27.128.225}  & $\bullet$  &   &   \\
\textbf{8.134.69.241}   & $\bullet$  &  &   \\
\textbf{47.105.167.35}  & $\bullet$  &  & \\
\textbf{api.djz8888.com}  &  & $\bullet$ &  \\
\textbf{sy.7ex3i.com}   & $\bullet$ & &  \\
\textbf{8.129.178.22:54321} & & $\bullet$ &  \\
\textbf{pay.namaye.cn} &  & $\bullet$ &  \\
\textbf{81.71.77.222:54321}  &  & & $\bullet$ \\
\textbf{rdvjjv.com:8666}  & $\bullet$ &  &       \\
\textbf{go.huabeitong.net}   & $\bullet$ &  &  \\
\textbf{39.108.233.232}  & $\bullet$ & &  \\
\textbf{shop.xingkongpay.com} &  & $\bullet$  &  \\
\textbf{pay.usdtpay1.com}  &  &  & $\bullet$  \\
\textbf{d49byu.mango100.xyz:2053}   & $\bullet$   &  & \\
\textbf{pay.upay.games} &  &  & $\bullet$ \\
\textbf{eic.keji668.com}  & $\bullet$ & & \\
\textbf{pay.it.10086.cn}  &  $\bullet$ &  &  \\
\textbf{120.24.26.22} &  $\bullet$ & &  \\
\textbf{wappay.189.cn} & $\bullet$  & &  \\
\textbf{t.taobaob.cc} & $\bullet$  & &  \\
\textbf{47.119.169.21:8124} & $\bullet$ & & \\
\bottomrule 
\end{tabular}
\label{tab:fourth_party_pay}
\end{table}